\documentclass[preprint,authoryear,5p]{elsarticle}  
\usepackage{graphicx}
\usepackage{amsmath}
\usepackage{amssymb}    
\usepackage{epsfig}
\usepackage[latin1]{inputenc}
\usepackage{hyperref,url}
\usepackage{placeins}
\usepackage{xcolor}

\usepackage[T1]{fontenc}
\usepackage{ae,aecompl,url}
\usepackage{multicol}
\newcommand\NS{N\!S}

\title{Excitation of Tumbling in Phobos and Deimos}

\author[a1]{Alice C. Quillen\corref{cor1}}
\ead{alice.quillen@rochester.edu}
\author[a1,a2]{Mckenzie Lane}
\ead{mlane39@gatech.edu}
\author[a1,a3]{Miki Nakajima}
\ead{mnakajima@rochester.edu}
\author[a1]{Esteban Wright}
\ead{ewrig15@ur.rochester.edu}

\address[a1]{Department of Physics and Astronomy, University of Rochester, Rochester, NY 14627 USA}
\address[a2]{School of Earth and Atmospheric Sciences, Georgia Institute of Technology, Atlanta, GA 30332 USA}
\address[a3]{Department of Earth and Environmental Sciences, University of Rochester, Rochester, NY 14627, USA}
\cortext[cor1]{Corresponding author}

\hypersetup{draft}

\begin{document}
\begin{abstract}
Mass-spring model simulations are used to investigate past spin states of a viscoelastic Phobos and Deimos.  From an initially tidally locked state, we find crossing of a spin-orbit resonance with Mars or a mean motion resonance with each other does not excite tumbling in Phobos or Deimos.  However, once tumbling  our simulations show that these moons can remain so for an extended period and during this time their orbital eccentricity can be substantially reduced.  We attribute the tendency for simulations of an initially tumbling viscoelastic body to drop into spin-synchronous state at very low eccentricity to the insensitivity of the tumbling chaotic zone volume to eccentricity.  After a tumbling body enters the spin synchronous resonance, it can exhibit long lived non-principal axis rotation and this too can prolong the period of time with enhanced tidally generated energy dissipation.  The low orbital eccentricities of Phobos and Deimos could in part be due to spin excitation by nearly catastrophic impacts  rather than tidal evolution following orbital resonance excitation.  
\end{abstract}


\maketitle

\section{Introduction}


Phobos orbits within Mars' geosynchronous orbit, and it is currently drifting inward due to tides excited by Phobos that are dissipated in Mars \citep{burns92,bills05}.  Tides excited by Mars that are dissipated in Phobos (satellite tides) increase Phobos' inward drift rate in semi-major axis 
and reduce Phobos' orbital eccentricity.  If Phobos' eccentricity is only affected by tidal evolution, then Phobos must have had a significantly higher orbital eccentricity in the past 
\citep{singer66,lambeck79,cazenave80,yoder82}.  
Because the semi-major axis drift rate is increased by satellite tides,   
Phobos' orbit would have crossed Deimos' orbit only 1 to 2 Gyrs ago \citep{cazenave80,szeto83}.
This would have caused orbital instability and collisions between the two moons \citep{szeto83}.
An epoch of high eccentricity and swift migration is not ruled out if Phobos and Deimos 
are remnants of earlier moons that collided (Efroimsky, private communication).
The tidal drift rates assumed by previous work (e.g., \citealt{cazenave80}) assumed
a tidally locked state for Phobos throughout tidal evolution.  However, the inferred previous high orbital eccentricities,  $e>0.2$, would have caused Phobos to tumble prior to entering
the spin synchronous state \citep{wisdom87}.  Tumbling would have elevated the energy dissipation rate in Phobos and the associated eccentricity damping rate by orders of magnitude \citep{wisdom87}, exacerbating the problem caused by Phobos' short tidal evolution timescale.


Alternatively, Phobos could have remained at low eccentricity and its current eccentricity value is due to excitation by other dynamical processes such as crossing spin-orbit resonances \citep{yoder82}. 
Phobos' current semi-major axis is  $a = 2.76 R_{\rm Mars}$, where $R_{\rm Mars}$ is
the mean equatorial radius of Mars.
When Phobos was at a semi-major axis of $a \approx 3.8 R_{\rm Mars}$, its orbital period would have been half that of Mars' spin rotation period.   This is a 2:1 spin-orbit resonance between
Phobos' orbit and Mars' rotation, with strength dependent on the quadrupole gravitational moment of Mars \citep{yoder82}, (also see appendix D by \citealt{frouard17}).
If initially low, Phobos' orbital eccentricity after passing through this spin-orbit resonance is estimated to be 0.032 and above its current value of 0.015 \citep{yoder82}.  
Because the satellite spin angular momentum is negligible compared to its orbital angular momentum,  
the rate of tidal energy dissipation $dE/dt$ is related to the rate of eccentricity change
$\dot e$ with the constraint that angular momentum is conserved.   
At low eccentricity
\begin{align} 
\frac{dE}{dt}  &\approx   \frac{1}{2} M n^2 a^2 \frac{de^2}{dt}
\end{align}
where $M$ is moon mass, and $n$ is its mean motion.
Phobos' current eccentricity value is consistent with an eccentricity excited previously while crossing this resonance and tidal eccentricity damping since then \citep{yoder82}.
If Phobos remained at low eccentricity during the past few Gyr then its semi-major axis drift rate would have remained slow enough to ensure its orbital stability and Deimos's orbit need never have been encountered.

Deimos' anomalously low eccentricity, $e=0.00033$, presents a similar problem. 
Inward tidal migration of Phobos would have increased Deimos' orbital eccentricity as they crossed the 2:1 orbital mean motion resonance where Deimos' orbital period is twice that of Phobos \citep{yoder82}.   
As the time for its excited orbital eccentricity to damp (in the tidally locked state) is greater than the age of the solar system,  Deimos would have retained a higher eccentricity than it currently has. 
This implies that there must be an additional mechanism damping or reducing Deimos' eccentricity.
\citet{wisdom87} speculated  ``Perhaps the resonance passage occurred much earlier when Deimos was more likely to still be chaotically tumbling. ..... The small orbital eccentricity of Deimos could also be a consequence of an episode of chaotic tumbling.'' 
Thus the low eccentricities of both Phobos and Deimos may be a result of prior episodes of tumbling.

The low orbital inclinations of Phobos and Deimos with respect to Mars' equatorial plane
suggest that they were born in or from a circumplanetary disk (e.g., \citealt{goldreich65,burns92,rosenblatt11,craddock11,rosenblatt16, HesselbrokMinton2017, CanupSalmon2018}), whereas an inferred previous higher orbital eccentricity  for Phobos has been used to argue for capture of Phobos  from heliocentric orbit \citep{burns92,lambeck79}.
Phobos and Deimos should remain within $1^\circ$ of their Laplace plane during tidal evolution \citep{cazenave80}, and this would be independent of Mars' obliquity variations as they take place adiabatically compared to Phobos' and Deimos' nodal precessional motion.
A low eccentricity history for Phobos and Deimos is concordant with their formation in and from a circumplanetary disk.

Both Phobos and Deimos are currently in a spin synchronous or tidally locked rotation states where their angular rotation rates match their mean motions.
However due to their elongation, Phobos and Deimos can chaotically tumble for long periods
of time even at their quite low orbital eccentricities \citep{wisdom87}.
\citet{wisdom87} 
speculated that prior to ending in a tidally locked or state,
``Phobos must have spent many hundreds of thousands, probably millions, of years in the chaotic zone''.   
\citet{wisdom87} estimated that the timescale for the damping of orbital 
eccentricity can be a few orders of magnitude shorter for a chaotically tumbling satellite than for a synchronously rotating satellite.  However, due to resonant substructure in the chaotic region and its dependence on attitude,  it's difficult to estimate when or how an initially tumbling and tidally dissipating body would exit a tumbling state.  
Eventually, the rotation could remain close to one of the attitude-stable islands bordering the chaotic zone long enough that tidal dissipation would let the satellite exit the chaotic tumbling dynamical state  \citep{wisdom87}.

Impacts significantly affect the spins distributions of asteroids  \citep{harris79} and their rotation states \citep{henych13}, excepting the very largest ones, or those with diameters above 100 km, \citep{farinella92}.
Mars' moons lie well beneath this size limit, so their spins are likely to have been significantly perturbed by rare energetic impacts from asteroids crossing Mars' orbit or secondary impacts from crater ejecta leaving Mars.  A high fraction, 
50\% of rocky asteroids and satellites exhibit craters with diameters of 1 times their mean radius ($R_m$),  and the great majority have more than two craters with diameters greater than $R_m/2$ \citep{thomas99}. 
The impact that formed the Stickney crater on Phobos was probably 
energetic enough to knock Phobos out of its tidally locked state \citep{weidenschilling79,ramsley19}.
Phobos would have tumbled prior to reentering the tidally locked state in which it currently resides \citep{wisdom87}. 

In this paper we reconsider processes for spin excitation and evolution of non-spherical moons Phobos and Deimos.
With mass-spring model viscoelastic dynamical simulations, we simulate spin-down, spin-orbit resonance  and tumbling in orbit about host planet Mars.  Prior studies (e.g., \citealt{wisdom87}) have integrated the equations 
of motion for dissipationless rigid bodies. Our numerical study work differs from prior work as we use
a viscoelastic model for the spinning body.  In our simulations,   tidal deformation, torque and dissipation are
generated within the interior of the spinning body, directly coupling orbital and spin evolution.
Our simulations are described in section \ref{sec:sim}.
In section \ref{sec:res} we test the possibility that orbital resonance crossing could  have excited the spins of Mars' moons.
In section \ref{sec:wob} we explore spin evolution from an initial tumbling state. 
We look at eccentricities reached from the tumbling state during the transition to the tidally locked state  and tidal dissipation rates while tumbling. In section \ref{sec:coll} we discuss collisional excitation as a mechanism for tumbling excitation.
A summary and discussion follow. 

\subsection{Physical Quantities}
\label{sec:phys}

In Table \ref{tab:phobos_deimos} we list orbital and physical values  for Phobos and Deimos.
Ratios and quantities relevant for our numerical simulations are also listed.

Phobos and Deimos are both in spin synchronous or tidally locked spin states.
We estimate the time it could have taken  spherical bodies in similar orbits to reach this state.
The spin-down time $t_{\rm despin}$ of an orbiting spherical moon 
can be estimated from its moment of inertia, $I$ and initial spin $\omega$, giving 
$t_{\rm despin} \sim I\omega/T$ \citep{peale77}
(see Eq. (9) by \citealt{gladman96}).   The tidal torque is $T$.
The secular part of the semi-diurnal ($l = 2$) term in the Fourier expansion of the perturbing potential
from point mass $M_*$, (here Mars)  gives a tidally induced torque on a spherical body  of mass $M$,
 radius $R$ and semi-major axis $a$ (e.g., \citealt{kaula64,goldreich63,efroimsky13}).
\begin{equation}
 T = \frac{3}{2} \frac{G M_*^2}{a} \left( \frac{R}{a} \right)^5 k_{2} \sin \epsilon_2.
\end{equation}
Here $k_2 \sin \epsilon_2$ is frequency dependent and is known as a quality function. It is often approximated as $k_2/Q$  with energy dissipation factor $Q$ and Love number $k_2$ describing
the deformation and dissipation in mass $M$.  See \citet{efroimsky15} for discussion on the accuracy of this approximation.
This torque gives a time for a spherical body to spin down
\begin{equation}
t_{\rm despin} \approx P \frac{1}{15\pi} 
\left(\frac{M}{M_*}\right)^\frac{3}{2} 
\left(\frac{a}{R}\right)^\frac{9}{2} \frac{Q}{k_2} \label{eqn:t_despin}
\end{equation}
where $P$ is the spin period. 
For an incompressible homogeneous spherical elastic body, 
the Love number $k_2 \sim 0.038 e_g/\mu$ where $\mu$ is the moon's rigidity and 
\begin{equation}
e_g \equiv \frac{GM^2}{R^4} \label{eqn:eg}
\end{equation}
is a measure of the moon's gravitational energy density
(following Equation 6 by \citealt{quillen17_pluto}).
Wobble damping timescales estimated from asteroid light curves
give an estimate for asteroid viscoelastic  material properties  $\mu Q \sim 10^{11}$ Pa \citep{pravec14}.
For energy dissipation factor $Q=100$ this gives  a 
material with rigidity  $\mu = 1$ GPa, roughly consistent with ice or rocky rubble.
Spin down times with $\mu Q =10^{11}$ Pa and $M_* = M_{\rm Mars}$ are listed in Table \ref{tab:phobos_deimos} and are computed using semi-major axes,  masses and mean radii for Phobos and Deimos that also listed in this Table.
The despin times are shorter than those used by \citet{peale77} as he 
assumed a larger value for $\mu Q$.

A time for wobble or non-principal axis rotation to decay  (following \citealt{peale77,gladman96})
\begin{equation}
    t_{\rm wobble} \approx t_{\rm despin} \left(\frac{n}{\omega} \right)^4 
\end{equation}
where $n$ is the mean motion and $\omega$ the spin. 
For a low eccentricity tidally locked body $n =\omega$ and  the time for non-principal axis rotation to decay is the same as the spin down time.
If  tumbling is continuously excited in an elongated body then it may remain tumbling for far longer than the spin down time of Equation \ref{eqn:t_despin}.

We check estimates for eccentricity damping due to satellite tides
in Phobos and Deimos using a timescale 
\begin{equation}
t_{\rm edamp} \approx \frac{2}{21} \frac{M}{M_{\rm Mars}} \left(\frac{a}{R}\right)^5 \frac{Q}{k_2} n^{-1}
\end{equation}
based on the formula for eccentricity damping for a tidally locked body in eccentric orbit
(following \citealt{peale78,yoder82} and \citealt{M+D} Equation 4.198).
The resulting eccentricity damping timescales
for Phobos and Deimos are listed in Table \ref{tab:phobos_deimos}.
We confirm that Phobos would have tidally damped its eccentricity
but Deimos would not have, confirming previous estimates \citep{lambeck79,yoder82,szeto83}.

The classic spin-orbit dynamics problem is that of a rigid non-spherical body undergoing principal axis rotation that is orbiting a massive central point mass at zero obliquity.  
The equation of motion 
\begin{align}
\ddot x + \frac{\alpha^2}{2}\left( \frac{a}{r}\right)^3 \sin (2 x - 2 f) = 0  \label{eqn:so1}
\end{align}
\citep{wisdom87,celletti10}
with $a, r, f$  respectively, the semimajor axis, the orbital radius and the true anomaly.
Here time $t$ is in units of the mean motion and $x$ is the angle
between the body's long axis and the pericenter line.  
The asphericity parameter $\alpha \equiv \sqrt{3(B-A)/C}$ and
 $A<B<C$ are the body's principal axis moments of inertia. 
The ratio $(B-A)/C \approx (b_b/a_b)^2 - (c_b/a_b)^2$ in terms of body axis ratios where $a_b,b_b,c_b$ are body semi-major axis lengths.
Using the body axis ratios for Phobos and Deimos 
we compute asphericity parameters for each moon and these are listed in Table  \ref{tab:phobos_deimos}.   The rotational dynamics will be discussed further in
section \ref{sec:wob}.
The dissipative spin-orbit problem depends on
the unitless parameter 
\begin{equation}
K_d  \equiv 3  \frac{k_2}{\xi Q} \left( \frac{R}{a} \right)^3 \frac{M_*}{M} \label{eqn:Kd}
\end{equation}
\citep{correia04,celletti10,celletti14}.
The parameter $\xi$ depends on the moment of inertia, $C =\xi M R^2$.
With $\mu Q =10^{11}$ Pa  and $\xi\approx 0.4$ we include
in Table \ref{tab:phobos_deimos} estimates for $K_d$ for Phobos and Deimos.



\begin{table*}
{
\centering
\caption{Quantities for Phobos and Deimos}
        \label{tab:phobos_deimos}
\begin{tabular}{lllll} 
\hline
&&                                           Phobos & Deimos &\\
\hline
Dimensions   & $2a_b,2b_b,2c_b$ & $27 \times 22 \times  18$ km  & $15 \times  12.2 \times 11 $ km \\
Body axis ratios   & $(b_b/a_b, c_b/a_b)$ & (0.82,0.67) & (0.82,0.73) \\
Mean Radius  & $R$ & 11.2667 km & 6.2 km  \\
Mass  &  $M$ & 	$1.0659\times 10^{16}$ kg & $1.4762 \times 10^{15}$ kg  \\
Mass ratio & $ M_{\rm Mars}/M$ & $6.0 \times 10^7 $  &  $4.3 \times 10^8 $\\ 
Eccentricity  & $e$ & 0.0151 & 0.00033  \\ 
Semi-major axis  & $a$ & $2.76  R_{\rm Mars}$ &  $6.91  R_{\rm Mars}$ \\
Orbital period   & $P=\frac{2\pi}{\omega}$ &  0.3189 day &  1.263 day  \\
Gravitational timescale & $t_{g}$ &1555 s  & 1418 s  \\
Spin  & $\omega  t_g  $ & 0.32 & 0.09 \\
Surface rotational velocity & $\omega R$ & 2.5 m/s & 0.36 m/s \\
Semi-Major axis/Radius & $a/R$ & 830.3 & 3777.6   \\
Mars spin & $\omega_{\rm Mars} t_g$ & 0.10 & 0.11  \\
Radius ratio & $R_{\rm Mars}/R$ & 300.8 & 546.7  \\
Despin time  & $t_{\rm despin}$ &  3000 yr  &  $4\times 10^5$ yr\\
Eccentricity damping time  & $t_{\rm edamp}$ & $0.4 \times 10^9$ yr & $2 \times 10^{12}$ yr  \\
Asphericity parameter & $\alpha$   & 0.82 & 0.65 \\ 
Dissipation parameter & $K_d$  & $10^{-7}$ & $10^{-9}$ \\
\hline
&& Mars & \\
\hline
Mars Radius & $R_{\rm Mars}$ &3389.5  km \\
Mars Mass & $M_{\rm Mars}$ & $6.4171 \times 10^{23}$ kg \\
Mars Rotation period & $P_{\rm Mars}$ & 1.025957 day \\
Synchron.  semi-maj. axis & $ a_{\rm sync}$ & $6 R_{\rm Mars}$\\
Moon Mass ratio & $M_{\rm Phobos}/M_{\rm Deimos}$ & 7.2 \\
\hline
\end{tabular}
\\
Orbital periods are sidereal.  The gravitational time $t_g$ is defined in Equation \ref{eqn:tg}.
Mars' Love number and tidal dissipation factor are those measured by \citet{bills05}.
Despin and eccentricity damping times and $K_d$ are estimated with $\mu Q = 10^{11} $ Pa, see
section \ref{sec:phys}.
}
\end{table*}

\begin{figure}
\centering\includegraphics[width=2.5in]{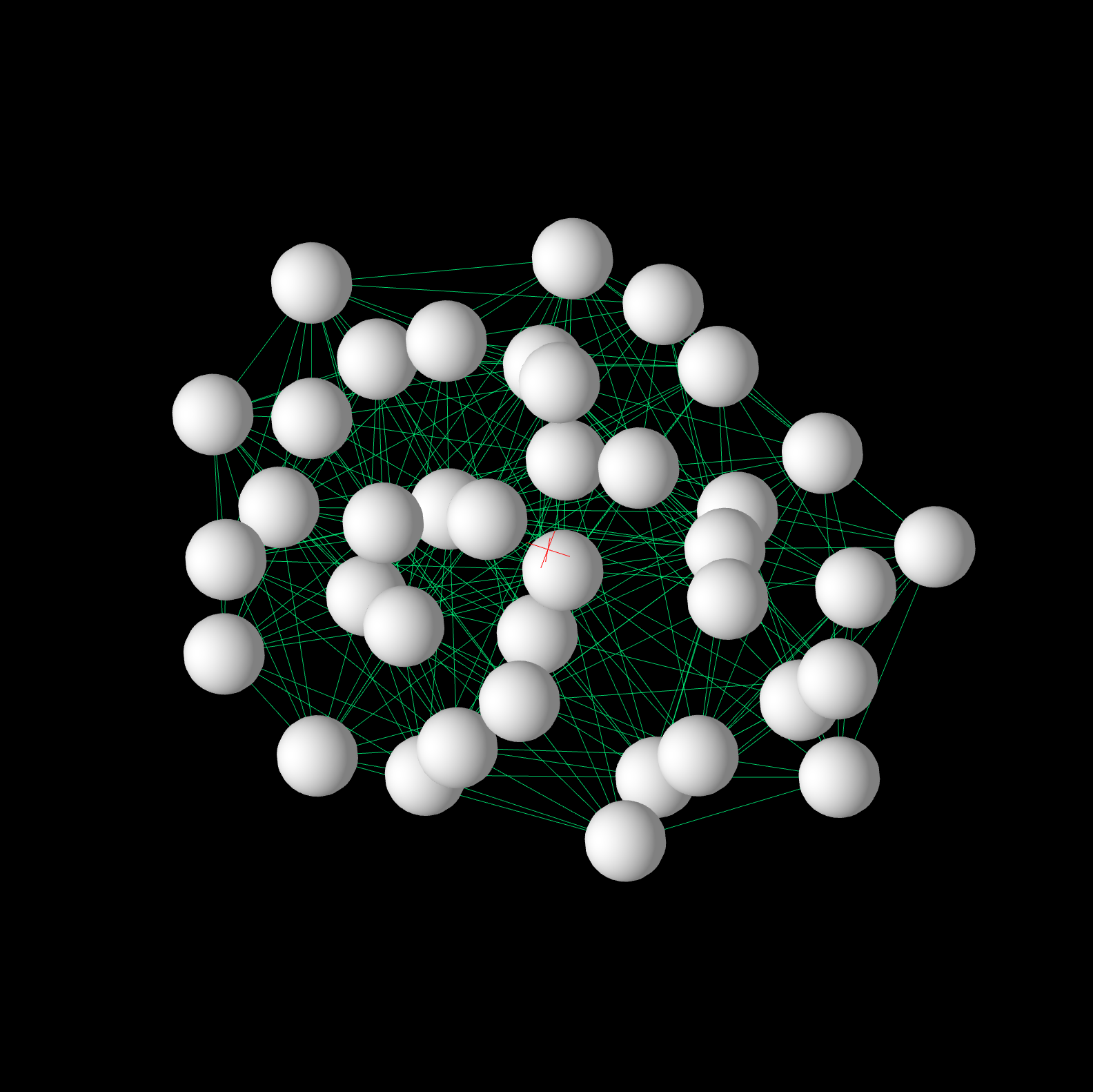} 
\caption{A snapshot of a simulation of Phobos showing the mass nodes and springs.
The image shows only the resolved moon.  The grey spheres represent the point masses
and the green lines are the connecting springs in the mass-spring network.
The diameter of the grey spheres is equal to half the minimum interparticle spacing.
\label{fig:snap}}
\end{figure}

\section{Numerical Simulations}
\label{sec:sim}

To explore tidal and spin evolution in migrating multiple planet and moon systems we use our mass-spring and N-body code \citep{quillen16_crust,frouard16}. Viscoelastic response arises naturally through dissipation in the springs and this makes it possible to explore scenarios where orbital migration and tidal evolution occur \citep{quillen17_pluto}. 
The mass-spring model is a viscoelastic and N-body code of low computational complexity that can accurately measure small deformations and conserves angular momentum \citep{quillen16_crust,frouard16,quillen16_haumea,quillen17_pluto,quillen19_bennu,quillen19_moon,quillen19_wobble}. 
Excited normal modes, excited non-principal axis rotation or tumbling,  libration, attitude instability, spin-orbit, spin-binary, spin-secular resonance and spin-precession mean-motion capture, all occur naturally within our simulations.   
The code accuracy (in its ability to measure small strains), simplicity and speed, compared to more computationally intensive grid-based or finite element methods, are strengths of this type of code.
The code has been used to measure tidal spin down \citep{frouard16,quillen16_haumea}, directly connecting numerically measured torques to the underlying viscoelastic model and without any the underlying analytic assumptions commonly used to compute tidal drift rates. The mismatch between predicted and numerically measured spin down rates \citep{frouard16} is not due to inaccuracy of the code, but due to neglect of bulk viscosity in the analytic computation. 
We have used the model to explore resonant obliquity evolution in long term integrations of rapidly spinning and non-round satellites \citep{quillen17_pluto} and with shorter integrations we resolve the internal tidal heating distribution in a spin synchronous Moon in eccentric orbit \citep{quillen19_moon}.  Comparison between numerically measured and analytically predicted wobble (or non-principal axis) damping rates illustrate that the code is accurate and that we understand how the simulation dissipation rates depend on the simulated rheology \citep{quillen19_wobble}.  

\begin{table}
\begin{center}
\caption{\large  Base Simulation Parameters \label{tab:common}}
\begin{tabular}{@{}lllllll}
\hline
Minimum interparticle distance & $d_I$ &  0.5 \\
Max spring length & $d_S$  & 2.3 $d_I$\\
Number of nodes                     & $N$  & $ 40$ \\
Number of springs  per node   & $\NS/N$ & $ 8$ \\
Spring constant          & $k_s$ & $0.22$\\
Young's modulus  &  $E$ & $\sim 2 $ \\
Time step & $dt$ & 0.001 \\
Damping time &  $t_{\rm damp}$ &100  \\
\hline
\end{tabular}
\end{center}
{Notes: Due to variations in the generation of the random spring network, the number of springs per node and number of nodes  varies from simulation to simulation.
$N$ varies by about $\pm 2$ and $\NS$ varies by about $\pm 20$.
Additional simulation parameters are listed in Tables \ref{tab:res} and \ref{tab:ewob}.
}
\end{table}

\subsection{Sizes and units}

Our simulations work with mass $M$ in units of the the mass of a single resolved spinning body.  
Distances in units of volumetric radius, $R = R_{\rm vol}$,  the radius of a spherical body with the same volume. For our mass-spring simulations we work with time in units of a gravitational timescale
\begin{align}
t_g  &\equiv \sqrt{\frac{R^3}{GM}} = \sqrt{ \frac{3}{4 \pi G \rho}   }  
\label{eqn:tg}
\end{align}
Gravitational timescales for Phobos and Deimos are listed in Table \ref{tab:phobos_deimos}.
Their low mean density causes these times to be longer than for denser objects.

%

\subsection{Simulations of Phobos and Deimos}

As done previously \citep{quillen17_pluto}, we consider two or three masses in orbit under the influence of gravity.
One mass is the non-spherical spinning body, here Phobos or Deimos, and it is resolved with masses and springs. 
A second mass is the central host planet, here Mars.  
In some of our simulations a third point mass is included that represents
the other moon.

Due to the proximity of Mars to its moons, the rotation of its non-spherical figure can cause
significant perturbations on its moon's orbits (see \citealt{yoder82}).   We model the static gravitational
field of Mars by taking into account the quadrupole terms in its gravitational potential,
\begin{align}
U(r,\phi,\psi) & = -\frac{GM_{\rm Mars}}{r} \bigg( 1+ 
 \sum_{l=2}^{L} \left( \frac{R}{r} \right)^l  \sum_{m=0}^{l}  \\
& \qquad \left(\tilde  C_{lm} \cos (m \phi )+ \tilde  S_{lm} \sin (m \phi) \right) P_{l}^m(\sin \psi) \bigg). \nonumber
\end{align}
The normalized (or dimensionless) coefficients from recent measurements of Mars' gravity field
\citep{genova16} are
\begin{align} 
\begin{array}{lrll}
\tilde C_{20,M}= &   -\texttt{8.75021132354E-04} \pm \texttt{1.252311E-11} \\
\tilde C_{22,M}= &   -\texttt{8.46359038694E-05} \pm \texttt{2.411019E-12} \\
\tilde S_{22,M}= &    \texttt{4.89346258602E-05}  \pm \texttt{2.415168E-12}
\end{array}.
\label{eqn:C22}
\end{align}
Here $\phi$ is longitude and $\psi$ is latitude on
Mars, not the colatitude $\theta$ that is often used for spherical coordinates.
The associated Legendre functions $P_2^2(x) = 3(1-x^2)$
and $P_2^0(x) = (3x^2-1)/2$.
The $\tilde C_{22,M}$ and $\tilde S_{22,M}$ coefficients of Mars sum in quadrature to an amplitude 
$\tilde c_{22} \approx 10^{-4}$.
    
To take into account the rotation of Mars' gravitational field we rotate the quadrupole components
 with an angular rotation rate $\omega_{\rm Mars}$.  The azimuthal angle with respect
 to a fixed direction (a true longitude)
$\phi_{in} = \phi + \omega_{\rm Mars} t$.  
The gradient of the quadrupolar gravitational potential 
is added as an additional force onto each point mass and mass node in the simulation,
 and equally and oppositely to the mass representing Mars.

As we only resolve Phobos or Deimos with the mass-spring model, tidal evolution due to tidal dissipation in Mars is neglected.  
To mimic inward or outward drift in orbital semi-major  axis (migration) due to tidal dissipation in Mars, we apply small velocity kicks to the resolved body using the recipe for migration given in Equations 8--11 by \citet{beauge06} and as we did previously for the migrating Pluto system (\citealt{quillen17_pluto}). 
The migration rate, $\dot a$, depends on an exponential timescale
$\tau_a$ (the parameter $A$ in  Equation 9 by \citealt{beauge06}), giving 
 $\dot a \sim a \tau_a^{-1}$.
We adopt a convention $\tau_a <0$ corresponding to inward migration
for Phobos as it lies within Mars' synchronous orbit.

For our random spring model, particle or node positions for the resolved body are drawn from an isotropic uniform distribution but only accepted into the spring network if they are within the surface bounding a triaxial ellipsoid,
$x^2/a_b^2 + y^2/b_b^2 + z^2/c_b^2 = 1$, and if they are more distant than $d_I$ from every other previously generated particle. 
Particles nodes that are not inside the ellipsoid are deleted.
For more discussion on the different particle models see \citet{quillen16_haumea}.
Here $a_b,b_b,c_b$ are the body's semi-major axes. We use a subscript to
differentiate between body semi-major axis and orbital semi-major axis.
Once the particle positions have been generated, every pair of particles within $d_S$ of each
other are connected with a single spring of spring constant $k_s$ and damping parameter $\gamma_s$.
Springs are initiated at their rest lengths.

Because we need to simulate the spin and orbit dynamics for hundreds of orbital periods, 
we use only a few mass nodes to resolve Phobos or Deimos, similar
to our study of the obliquity evolution of Pluto and Charon's minor satellites \citep{quillen17_pluto}.
The Young's modulus and viscoelastic relaxation timescale are estimated
from the spring parameters,  (described previously by \citealt{frouard16})
and here these quantities are approximate because of the sparse network.
Sensitivity  to the number of springs and nodes in the network was  discussed
by \citet{quillen16_haumea}.
A simulation snap shot is shown in Figure \ref{fig:snap} to illustrate the spring network.

Because of self-gravity the body is not exactly in equilibrium at the beginning of the simulation and the body initially vibrates.
We apply additional damping in the springs at the beginning of the simulation
for a time $t_{\rm damp}$ to remove these vibrations.  
Only after $t_{\rm damp}$ do we process simulation
outputs for measurements of dissipation.  After this time spring parameters are not varied.
The numbers of springs, nodes and masses remain fixed during the simulation. When initializing orbital motion, the mean anomalies, longitudes of the ascending node, and arguments of periapse are chosen using uniform random distributions.
Simulation outputs are separated by the time interval $\Delta t= 10$ in our N-body units.

Base parameters for most of the  simulations discussed here are listed in Table \ref{tab:common}. 


\begin{table}
\begin{center}
\caption{\large  Simulations for Resonance Crossings \label{tab:res}}
\begin{tabular}{@{}lllllll}
\hline
Simulation                   & Ph$_{M2:1}$ & De$_{P2:1}$  \\
\hline
Resolved body         & Phobos                 & Deimos   \\
Approximate Body axis ratios        & 0.8,0.7   & 0.8,0.7 \\
$M_{\rm Mars}/M$   & $6.0\times 10^7$  & $4.3 \times 10^8$ \\
$R_{\rm Mars}/R$   &  301                        & 547 \\
Spin $\omega_{\rm Mars} t_g$    & 0.10           & 0.11 \\
Quadrupole $c_{22}$               & $10^{-4}$ & $10^{-4} $ \\
Initial $a_{\rm Phobos}/R$   & 1200           & 2390    \\
Initial $a_{\rm Deimos}/R$   &                  & 3778    \\
Initial $a_{\rm Phobos}/R_{\rm Mars}$   & 4.0           & 4.7    \\
Initial $a_{\rm Deimos}/R_{\rm Mars}$   &     -          &    6.9    \\
$M_{\rm Phobos}/M_{\rm Deimos}$   &          & 700       \\
Inward drift rate   $|\tau_{a,{\rm Phobos}}|^{-1} t_g$   &  $2 \times 10^{-6}$   & $10^{-7}$  \\
Initial eccentricity Phobos   &  0.001       & 0.0 \\
Initial eccentricity Deimos   &     -          & 0.0 \\
Spring damping parameter $\gamma_s$ &  3 & 3 \\
Figures & \ref{fig:phobos21} & \ref{fig:deimos21} \\
\hline
\end{tabular}
\end{center}
{Notes:   The simulations are shown in the Figures listed at the bottom of the table.
Additional simulation parameters are listed in Table \ref{tab:common}.
The radius of Mars, its spin rate and the normalized quadrupole moment $c_{22}$
are used to compute the force due to Mars's rotating
quadrupolar moments.    The mass ratios of Mars to resolved body
are those of Mars and Phobos or Mars and Deimos.  
The radius ratios are also based on real values.
The spin of Mars  $\omega_{\rm Mars}$ in gravitational units
 is computed using the current angular rotation rate of Mars.
The mass and radius $M, R$ are those of the resolved body.
The Deimos simulation includes Phobos as a point mass.
The inward drift rate $\tau_a^{-1}$ is applied to Phobos which is the resolved body
in the Phobos simulation but is a point mass in the Deimos simulation.
The Phobos simulation does not include a point mass for Deimos.
The gravitational timescales, $t_g$, for Phobos and Deimos are computed
using the masses and radii in Table \ref{tab:phobos_deimos}.
Orbital inclinations are initially set to 0.  The De$_{P2:1}$ simulation
was run with $t_{\rm damp} = 10^4$ to aid in damping initial libration 
 in the spin synchronous state prior to resonance crossing.
}
\end{table}

\begin{figure}
\centering\includegraphics[width=3in]{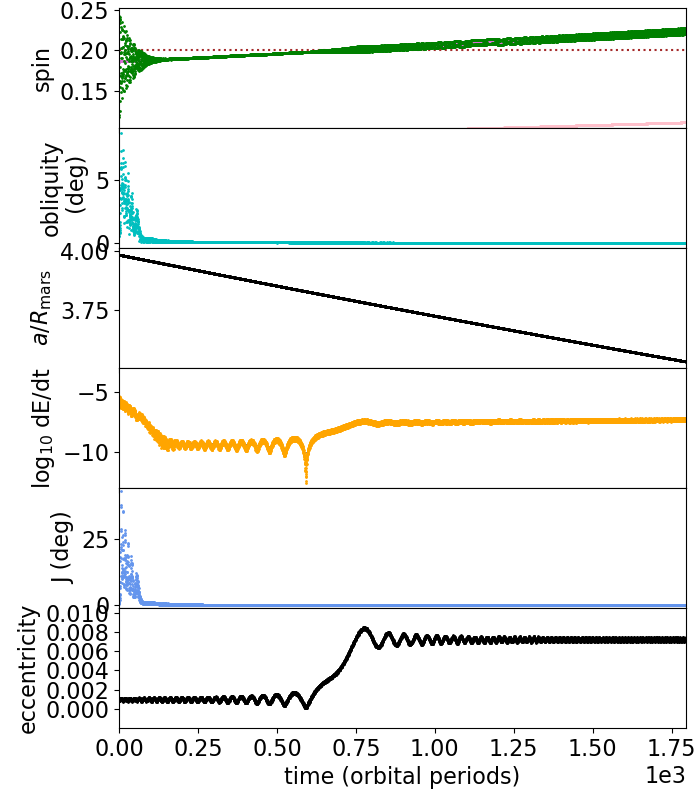} 
\caption{A simulation of a spinning Phobos drifting inward and crossing the 2:1 resonance with Mars'
figure and associated with its rotating equatorial gravitational quadrupolar moment.
Parameters for this simulation, denoted Ph$_{M2:1}$, are listed in Table \ref{tab:res}.
From top to bottom the panels show Phobos' spin, obliquity in degrees, semi-major axis in units of Mars' radius, energy dissipation rate (in Phobos), non-principal angle $J$ in degrees and orbital eccentricity.  
In the top panel, the green points show spin and the brown dotted line shows the location of the 2:1 orbital resonance with Mars' figure.
The $x$-axis is time in units of initial orbital period. 
This 2:1 resonance is crossed at $t \approx 0.75 \times 10^3$ orbital periods and increases the orbital eccentricity of Phobos from 0.001 to $\approx 0.008$.    
The energy dissipation rate increases at the same time due to the higher eccentricity.
After damping libration ($t<10^4$), the simulated Phobos remains tidally locked, at low obliquity and rotating about a principal body axis.   
}
\label{fig:phobos21}
\end{figure}

\section{Simulations of Resonance Crossings}
\label{sec:res}

In this section we test the hypothesis that crossing of an orbital resonance could
excite the spins of Phobos or Deimos.   Two orbital resonances were 
discussed by \citet{yoder82,wisdom87}, a 2:1 resonance between Phobos mean motion and Mars' spin 
and a 2:1 mean motion resonance between   Phobos and Deimos.  Phobos' eccentricity 
would have been increased by
the 2:1 resonance with Mars' rotation where Phobos' orbital period is half that of Mars' spin period.
Phobos would have crossed this resonance as it drifted inwards and 
when it was near an orbital semi-major axis of $3.8 R_{\rm Mars}$.
The other resonance is the 2:1 mean motion resonance between Deimos and Phobos
which would have been crossed when Phobos was near 
an orbital semi-major axis of $4.3 R_{\rm Mars}$.     This resonance
would have significantly increased the eccentricity of Deimos.

We show two simulations, one with a resolved Phobos, labelled Ph$_{M2:1}$  and
crossing the 2:1 resonance with Mars' rotation and the other with a resolved Deimos
and labelled De$_{P2:1}$ crossing the 2:1 mean motion resonance with Phobos.
Simulation parameters for these two simulations are listed in Table \ref{tab:res}.
Initial orbital inclinations are zero.
For both of these simulations Phobos is forced to migrate inward with a migration timescale $\tau_a$, and mimicking drift due to tidal dissipation in Mars.  
The mass of Mars in units of the spinning body is used for the central body.
The radius of Mars in units of the volumetric radius of the spinning body, 
and the spin rate of Mars in units of $t_g$ (for Phobos or for Deimos)
are used to compute the quadrupole force terms from Mars.  

Figure \ref{fig:phobos21} shows the simulation of Phobos crossing the 2:1 resonance with Mars' figure
and with parameters listed in Tables \ref{tab:common} and \ref{tab:res}.
In Figure \ref{fig:phobos21} we plot quantities as a function of time in units of 
the initial orbital period of the spinning moon (here Phobos).

In Figure \ref{fig:phobos21}
orbital elements for the spinning moon 
are computed using its center of mass position and velocity and the position, velocity and mass of Mars and neglecting its oblateness.  The angular momentum of the spinning body, ${\bf L}_s$, 
is computed at each simulation output by summing
the angular momentum of each particle node, using node positions and velocities
measured with respect to the center of
mass of the spinning body.  The moment of inertia matrix of the spinning body, $\bf I$,  in the fixed reference frame, is similarly computed from node positions and masses.  The moment of inertia
matrix is diagonalized and the eigenvector corresponding to the largest
eigenvalue is identified as the body's principal axis.  

The instantaneous spin vector $\boldsymbol \omega$ is computed by multiplying the spin angular momentum vector by the inverse of the moment of inertia matrix, ${\boldsymbol \omega} = {\bf I}^{-1} {\bf L}_s$.
The spin $\omega = |{\boldsymbol \omega}|$ is the magnitude of this spin vector and shown in the top panel in Figure  \ref{fig:phobos21}  in units of $t_g^{-1}$. 
At each simulation output, the obliquity is computed from the angle between the body's spin vector and its orbit normal.
One of the Andoyer-Deprit variables, the non-principal angle, $J$, 
is computed from the angle between angular
momentum vector and the body's principal axis with the largest moment of inertia.
The energy dissipation rate is measured from the strain rates in
the springs  (see \citealt{quillen19_moon,quillen19_wobble}).  The energy dissipation
rate in the figure is in N-body units or $e_g/t_g$ (as defined in Equations \ref{eqn:eg} and \ref{eqn:tg}). Here, we only compare relative variations in the energy dissipation rates.   In studies with a larger number of nodes and springs we have quantitatively measured energy dissipation 
directly from the simulations \citep{frouard16,quillen19_wobble}. 

From top to bottom, the panels in Figure \ref{fig:phobos21} show the simulated
Phobos' spin, obliquity in degrees, semi-major axis in units of Mars' radius, energy dissipation rate (in Phobos), non-principal angle in degrees and orbital eccentricity.
In the top panel, the green points show spin and the brown dotted line shows
the location of the 2:1 orbital resonance with Mars' figure.   The resonance is
crossed where the green points cross the brown dotted line.
The 2:1 resonance with Mars' rotation is crossed at $t \approx 0.75 \times 10^3$ orbital periods and increases the orbital eccentricity of Phobos from 0.001 to $\approx 0.008$.
The energy dissipation rate (in Phobos) is approximately 64 times higher after crossing the resonance ($\log_{10} (64) = 1.8$)
and this ratio is consistent with a tidal heating rate in the tidally locked state that is approximately proportional to $e^2$, as expected \citep{kaula64}. 
The libration caused by the increased orbital eccentricity 
is evident as a thickening in the breadth of the green points, showing spin, after the resonance is crossed, seen in the top panel of Figure \ref{fig:phobos21}.

The eccentricity jump (0.008) caused by the resonance in the Ph$_{M2:1}$ simulation is  smaller than
the 0.03 estimated by \citet{yoder82}.  We attribute our lower final eccentricity values to 
the drift rate, not to errors in the calculation by \citet{yoder82}.\footnote{The exponents should be 2 (not 12) for the coefficients in the top two lines in the left column of page 336 by \citet{yoder82} and his coefficient $A_{22}$ depends on the ratio $C_{22}/a$ not $C_{22}/A$ as written.} 
Using modern measurements of Mars, we agree with the sizes of the resonant coefficient $C_{22}$ and predicted eccentricity jump estimated by \citet{yoder82}.
By varying Phobos' drift rate $\tau_a$, 
we find that the eccentricity jump is smaller if Phobos is migrating more quickly.
The eccentricity jump is lower than predicted by \citet{yoder82} because our simulations are not 
drifting sufficiently slowly to be well inside the adiabatic regime (e.g., \citealt{quillen06}).
 The size of the eccentricity jump crossing this resonance can be estimated using Fresnel integrals, following appendix A by \citet{laskar04}.
Following equation 30 by \citet{yoder82}
\begin{equation}
\frac{de}{dt} \approx n \frac{3}{2} \left(\frac{R}{a} \right)^2  c_{22} \sin \phi
\end{equation}
where  the resonant angle  $\phi = \lambda - 2 \theta_{\rm Mars} - \varpi$, the angle $\lambda$ is Phobos'  mean longitude, $\varpi$ is Phobos' longitude of pericenter, and   $\theta_{\rm Mars}$
is a rotation orientation angle for Mars.
We approximate the time variation of the resonant angle as 
\begin{equation}
\phi \approx (n - 2 \omega_{\rm Mars} ) t + \dot n \frac{t^2}{2} + {\rm constant} ,
\end{equation} 
neglecting the resonant perturbation during resonant crossing.
The drift rate in Phobos' mean motion 
$\dot n  = \frac{3}{2} \frac{n}{\tau_a}$ in terms of our drift parameter.
Neglecting the sensitivity to initial phase,
Equation A.8 by \citet{laskar04}  gives an eccentricity jump
\begin{equation}
\Delta e \approx \sqrt{\frac{4 \pi }{3} \tau_a n }\ \frac{3}{2} \left(\frac{R}{a} \right)^2 c_{22} .
\end{equation}
Using the drift rate and mean motion from our simulation,
we computed $\tau_a n = 9 \times 10^4$.  This,  the quadrupolar moment and the above
equation gives  
 $\Delta e \sim 0.006$ which is  close to that observed in our simulation
(0.008).

After an initial spin down period (at $t<10^4$), Phobos remains tidally locked, at low obliquity and rotating about a principal body axis.    Passage through the resonance does increase the energy dissipation rate as the orbital eccentricity increases, but not as much as it would have if the body had started tumbling.   Because the body spins down into a tidally locked state prior to encountering the orbital resonance, the final spin state is insensitive to initial conditions.   However there is some variation between different simulations in the size
of the eccentricity jump that takes place when the resonance is crossed,  as it depends on the phase of the resonant angle when the resonance
is encountered \citep{yoder79,quillen06}.
A similar simulation but with a three times larger $c_{22}$ coefficient gives a higher eccentricity jump and also fails to push the simulated Phobos out of
the tidally locked state.  We conclude that crossing the 2:1 orbital resonance with Mars' figure rotation was unlikely to push Phobos out of its tidally locked state or cause it to tumble. 

\begin{figure}
\centering\includegraphics[width=3in]{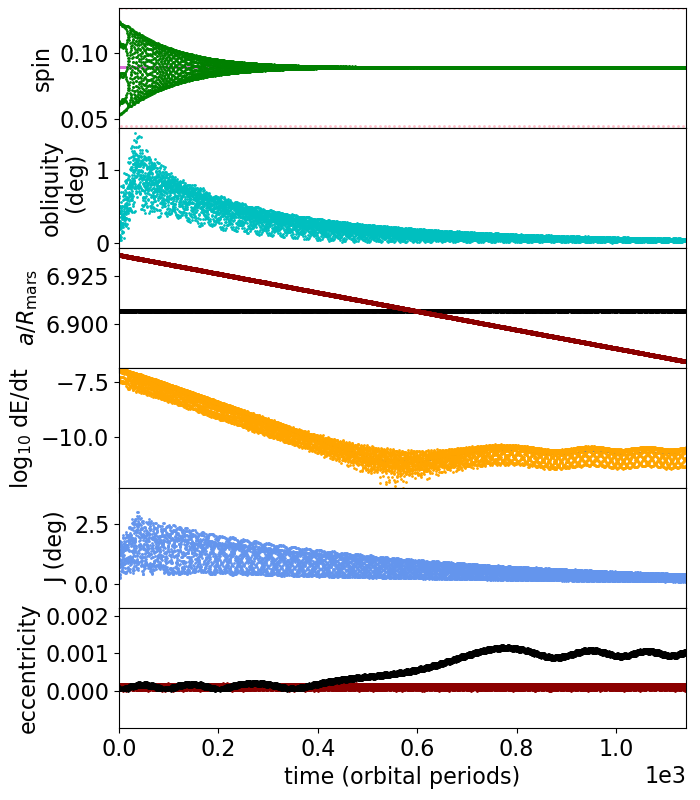} 
\caption{A simulation of a spinning Deimos crossing the 2:1 mean motion resonance with an inward drifting Phobos.  Panels are similar to those in Figure \ref{fig:phobos21} but this shows the De$_{P2:1}$ simulation with parameters listed in Table \ref{tab:res}.
In the third panel from top the semi-major axis times $2^{-\frac{2}{3}}$ of Phobos is shown in brown (the inclined line) and the semi-major axis of Deimos is shown in black (the horizontal line).  The 2:1 mean motion resonance occurs where these two lines cross. In the bottom panel the eccentricities for Deimos and Phobos are shown 
in black and dark red (remaining low), respectively.  Phobos is modeled as an inward
drifting point mass and we have enhanced its mass by a factor of 100.
Crossing the mean motion resonance increases the orbital eccentricity of Deimos
but does not significantly excite its spin. 
\label{fig:deimos21}
}
\end{figure}

In Figure \ref{fig:deimos21} we show a simulation, denoted De$_{P2:1}$, where the spinning body represents Deimos.  
This simulation is designed to see how the 2:1 mean motion resonance between Phobos and Deimos would affect the spin state of Deimos.
As Deimos is near Mars' synchronous radius (its mean motion is similar to the rotation rate of Mars), Deimos' semi-major axis drifts slowly due to tides excited in Mars.
It is Phobos' inward migration that  lets the two moons cross this mean motion resonance.
Due to the low mass ratio $m_{\rm Phobos}/m_{\rm Mars} \sim 10^{-8}$, the resonance is weak.  To see the effect of the resonance, Phobos must drift slowly compared to the resonance libration time (e.g., \citealt{quillen06}). 
To reduce the time required for the simulation, we increased the mass
of Phobos by a factor of 100 relative to its actual mass ratio with Mars.

Figure \ref{fig:deimos21} of a spinning Deimos is similar to Figure \ref{fig:phobos21} for  Phobos, and its parameters are also listed in Table \ref{tab:res}.
In this simulation, Phobos is included as a third point mass and is again forced to migration inward.
In the third panel from top, the semi-major axis times $2^{-\frac{2}{3}}$  of Phobos is shown in brown and the semi-major axis of Deimos is shown in black.  
The 2:1 mean motion resonance occurs where these two lines cross. In the bottom panel the eccentricities for Deimos and Phobos are shown in black and dark red, respectively.  
The ratio of Phobos' to Deimos' mass is 100 times larger than their actual mass ratio, to reduce the drift rate needed to see the effect of this resonance. 
The resonance does increase the eccentricity of Deimos, but slowly.  
Even with a much higher Phobos mass, crossing the resonance does not excite Deimos' spin. 

Using scaling for first order resonances \citep{quillen06,mustill11}, we can check to see if  Phobos' semi-major axis drift rate  would have been sufficiently slow
that the 2:1 mean motion resonance between Phobos and Deimos resonance was crossed adiabatically.   
The resonance is crossed adiabatically if the drift rate in Phobos' mean motion $\dot n/n^{2}$ is significantly smaller than  the mass ratio 
$(m_{\rm Phobos}/m_{\rm Mars})^\frac{4}{3}   = 4.2\times 10^{-11}$ \citep{quillen06,mustill11}.
Using Phobos' current semi-major axis drift rate $\dot n/n^2 = 5 \times 10^{-13}$
\citep{bills05} and scaling to the semi-major axis of the resonance at $a = 4.3 R_{\rm Mars}$, 
we estimate that Phobos' drift rate
would have been about 1000 times slower than the critical value defining the adiabatic limit for this resonance.
Thus Deimos' low eccentricity is not because the 2:1 mean motion resonance was crossed too quickly to increase Deimos' eccentricity.  


In both simulations, crossing orbital resonances due to an inward drifting Phobos
did not excite the spin of a tidally locked Deimos or Phobos.  
We rule out the possibility that crossing the strongest 
orbital resonances also excited  Phobos' or Deimos' spin.

\section{Evolution of  tumbling bodies}
\label{sec:wob}

\begin{table}
\begin{center}
\caption{\large  Simulations of Tumbling \label{tab:ewob}}
\begin{tabular}{@{}lllllll}
\hline
\multicolumn{3}{l}{Mass ratio $M_*/M$}            & \multicolumn{2}{l}{$10^4$}     \\
\multicolumn{3}{l}{Initial semi major axis $a/R$} & \multicolumn{2}{l}{48}       \\
\multicolumn{3}{l}{Initial mean motion $n t_g$} & \multicolumn{2}{l}{0.30}   \\
\multicolumn{3}{l}{Orbital period        $P= 2 \pi/(n t_g)$}   & \multicolumn{2}{l}{20.9}    \\
\multicolumn{3}{l}{Number of simulations in each series} & \multicolumn{2}{l}{10}  \\
\hline
Resolved body       &  \multicolumn{2}{l}{Deimos}     &      \multicolumn{2}{l}{Phobos}     \\
Simulation   series            & De$_F$  &   De$_S$  &  Ph$_F$ & Ph$_S$\\
Asphericity  $\alpha$ &    0.7 & 0.7 & 0.85& 0.85  \\
Initial eccentricity    &  0.2  & 0.05  & 0.2& 0.05 \\ 
Spring damping  $\gamma_s$ &  0.1& 0.01 & 0.1 &0.01\\
Figures & \ref{fig:De_all}a,\ref{fig:Ee_De}a & \ref{fig:De_all}b,\ref{fig:Ee_De}b 
& \ref{fig:Ph_all}a,\ref{fig:Ee_Ph}a & \ref{fig:Ph_all}b,\ref{fig:Ee_Ph}b,\ref{fig:NPA}\\
\hline
\end{tabular}
\end{center}
{Notes:  These simulations are shown in the Figures listed at the bottom of the table.
Additional simulation parameters are listed in Table \ref{tab:common}.
Orbital inclinations are initially set to 0.  The forced drift rates are zero,
$\tau_a^{-1} = 0$ and the quadrupole coefficient $c_{22}=0$.
Mean motions and orbital periods are given in gravitational units.
Simulation in each series have the same node and spring networks but initial conditions
were slightly different in initial obliquity, mean anomaly and spin.
Approximate body axis ratios are 0.8 and 0.7.
}
\end{table}

\begin{figure*}
\includegraphics[trim=0 0 0 0,clip,width=3.2truein]{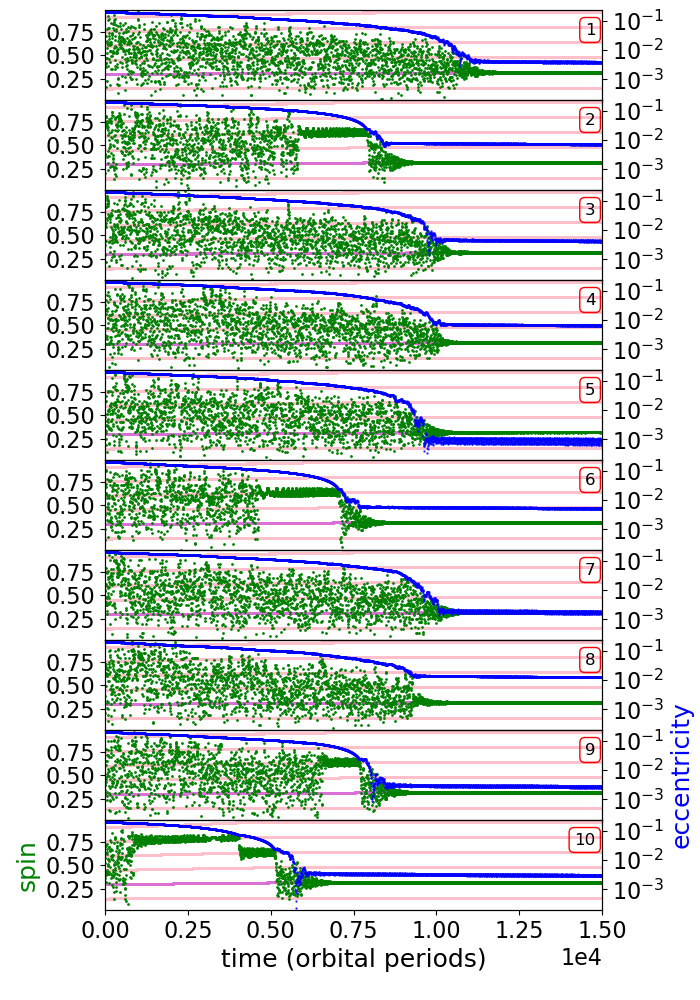}  
\includegraphics[trim=0 0 0 0,clip,width=3.2truein]{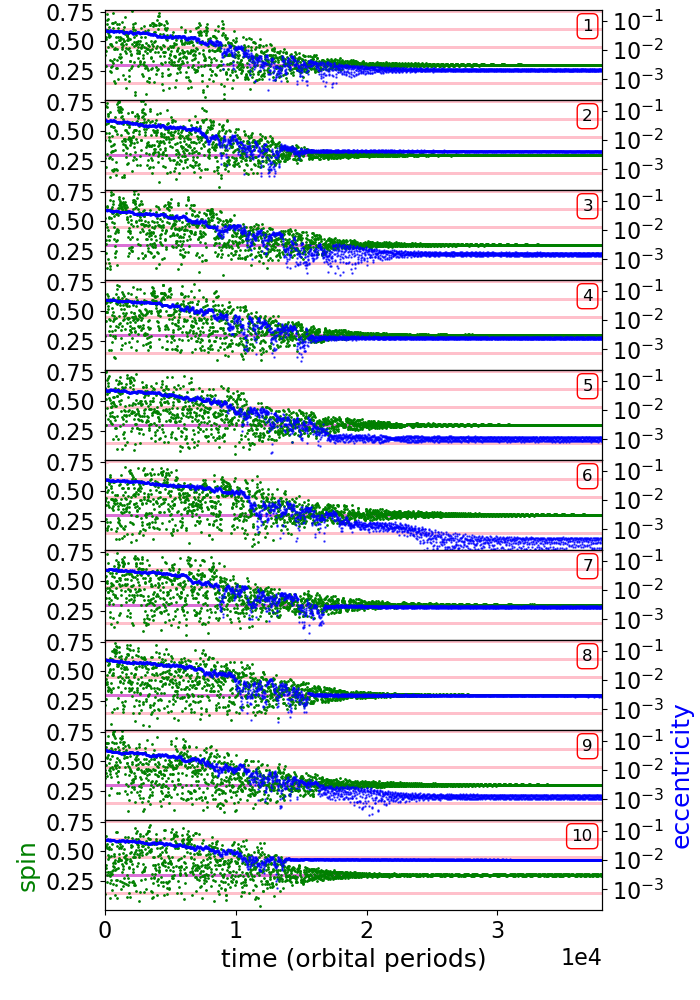} 
\caption{Spin and eccentricity evolution for series of Deimos simulations with parameters listed in Table  \ref{tab:ewob}. a) (left panels) Each panel shows a single
simulation from the De$_F$  series.
b) (right panels) The  De$_S$ simulation series.
The spring damping rate and initial eccentricity are higher
in the De$_F$  simulations than in the De$_S$ simulations. In each panel the green points show spin with axes on the left.  Blue points show eccentricity on a log scale and with axes on the right.  The x-axis shows time in orbital periods. The pink lines show the location of the spin-orbit resonances, with the dark pink lower one the spin-synchronous state.  As the eccentricity drops, the body eventually drops into the spin-synchronous state.  For reference, individual simulations  are numbered in the red boxes on the right.  The spinning body can be captured into the 2:1 spin-orbit resonance, as seen in the second from top simulation on the left between $t = 0.5$ and 0.75 orbital periods and in intervals in some of the other simulations from the De$_F$ series.
\label{fig:De_all}} 
\end{figure*}

\begin{figure*}
\includegraphics[trim=0 0 0 0,clip,width=3.2truein]{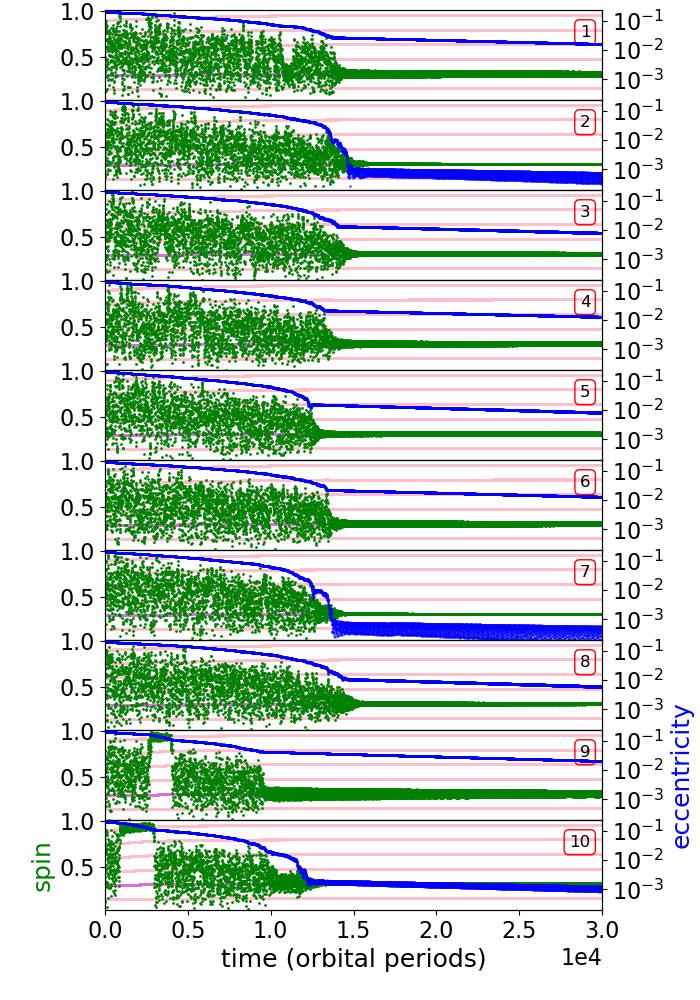}  
\includegraphics[trim=0 0 0 0,clip,width=3.2truein]{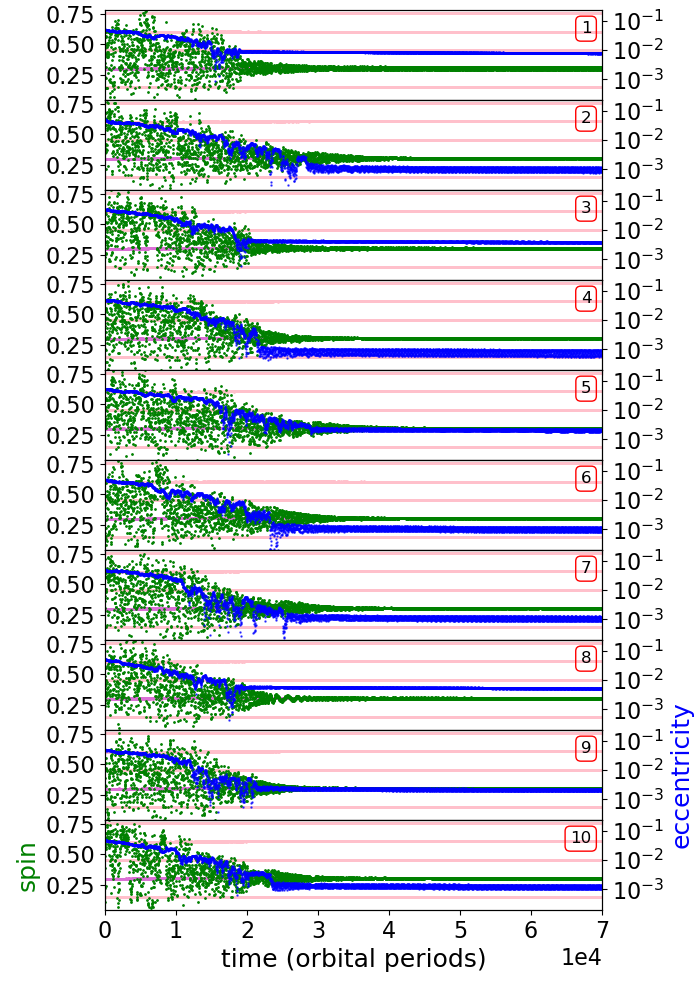}  
\caption{Spin and eccentricity evolution for series of Phobos simulations with parameters listed in Table  \ref{tab:ewob}.  Similar to Figure \ref{fig:De_all}.
a) (left panels) Each panel shows a single
simulation from the Ph$_F$  series.  
b) (right panels) The  Ph$_S$ simulation series. 
The spring damping rate and initial eccentricity are higher
in the Ph$_F$  simulations than in the Ph$_S$ simulations
\label{fig:Ph_all}} 
\end{figure*}

\begin{figure*}
\includegraphics[trim=0 0 0 0,clip,width=3.2truein]{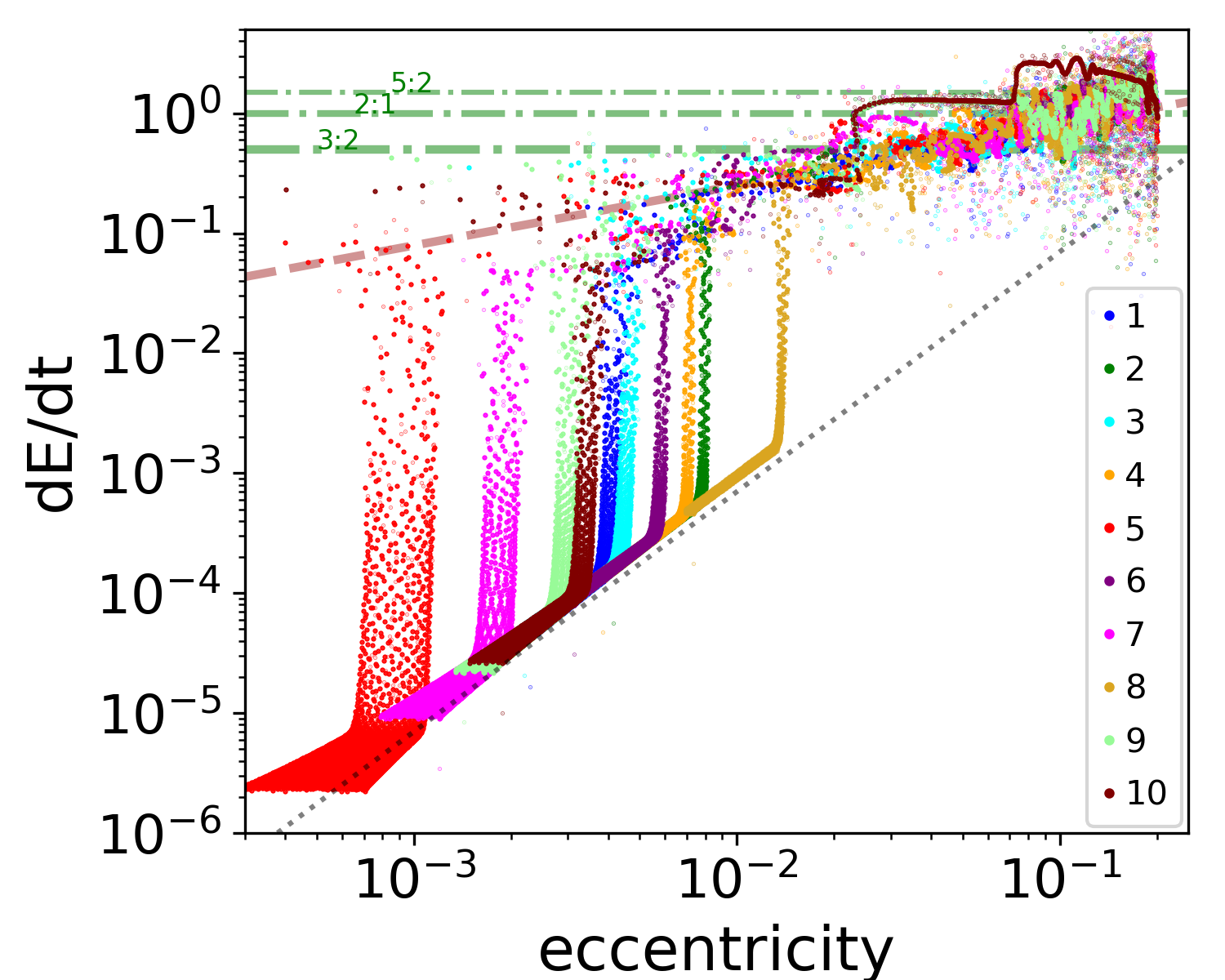}  
\includegraphics[trim=0 0 0 0,clip,width=3.2truein]{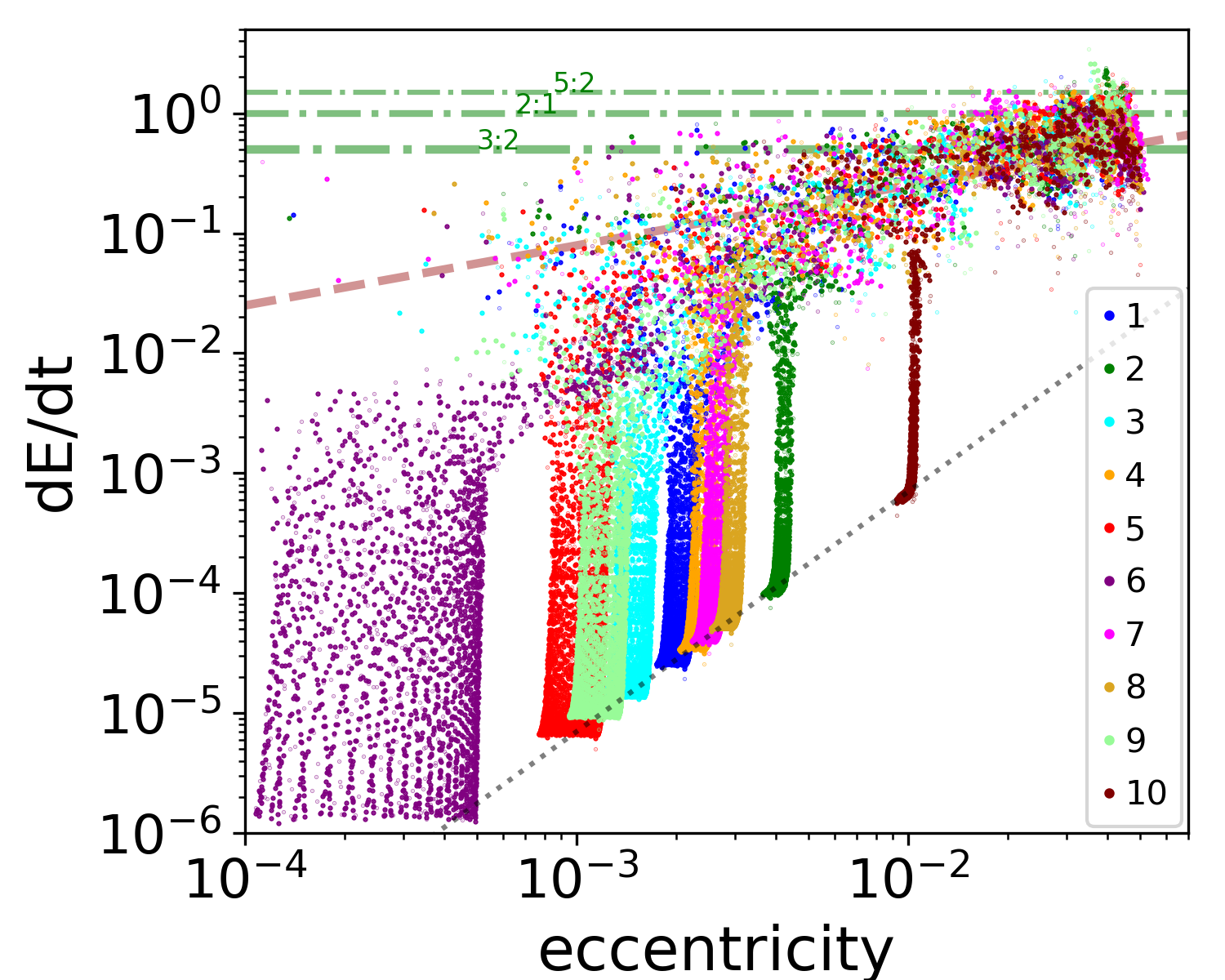}  
\caption{Energy dissipation rate vs eccentricity for the simulations shown in Figure \ref{fig:De_all}.  
a) (on the left) The De$_F$ simulations.
b) (on right) The De$_S$ simulations.
Each individual simulation is shown with different color points.  
The numbers in the key refer to the simulation number shown 
 in the red boxes in Figures \ref{fig:De_all}a or b. 
In each simulation the body starts in a tumbling  state and with a high energy dissipation rate.   
The dissipation rate drops when the body enters a spin synchronous state.
Simulation trajectories start on the right at higher eccentricity and move to the left as tidal dissipation causes the orbital eccentricity to drop. 
The axes are on a log scale.
Energy dissipation rates in each panel are normalized so that the dissipation rate in the
2:1 spin-orbit resonance should have $dE/dt = 1$.    A few of the De$_F$ simulations spend intervals of time in this 2:1 spin-orbit resonance.  
The grey dotted lines show estimates for the dissipation rate for low obliquity and tidally locked states for bodies undergoing principal axis rotation and with slope given by $dE/dt \propto e^2$.
The thick brown dashed line shows an estimate for the energy dissipation rate in tumbling states.  
The green lines are estimates for energy dissipation in 3:2, 2:1 and 5:2 spin-orbit resonances.
These simulations show that very low eccentricity can be reached from  tumbling 
rotation states.  As expected, the energy dissipation rate is  many orders of magnitude higher 
in a tumbling state or in a spin-orbit resonance than in tidally
locked states at low eccentricity, as previously emphasized by \citet{wisdom87}.
\label{fig:Ee_De}} 
\end{figure*}

\begin{figure*}
\includegraphics[trim=0 0 0 0,clip,width=3.2truein]{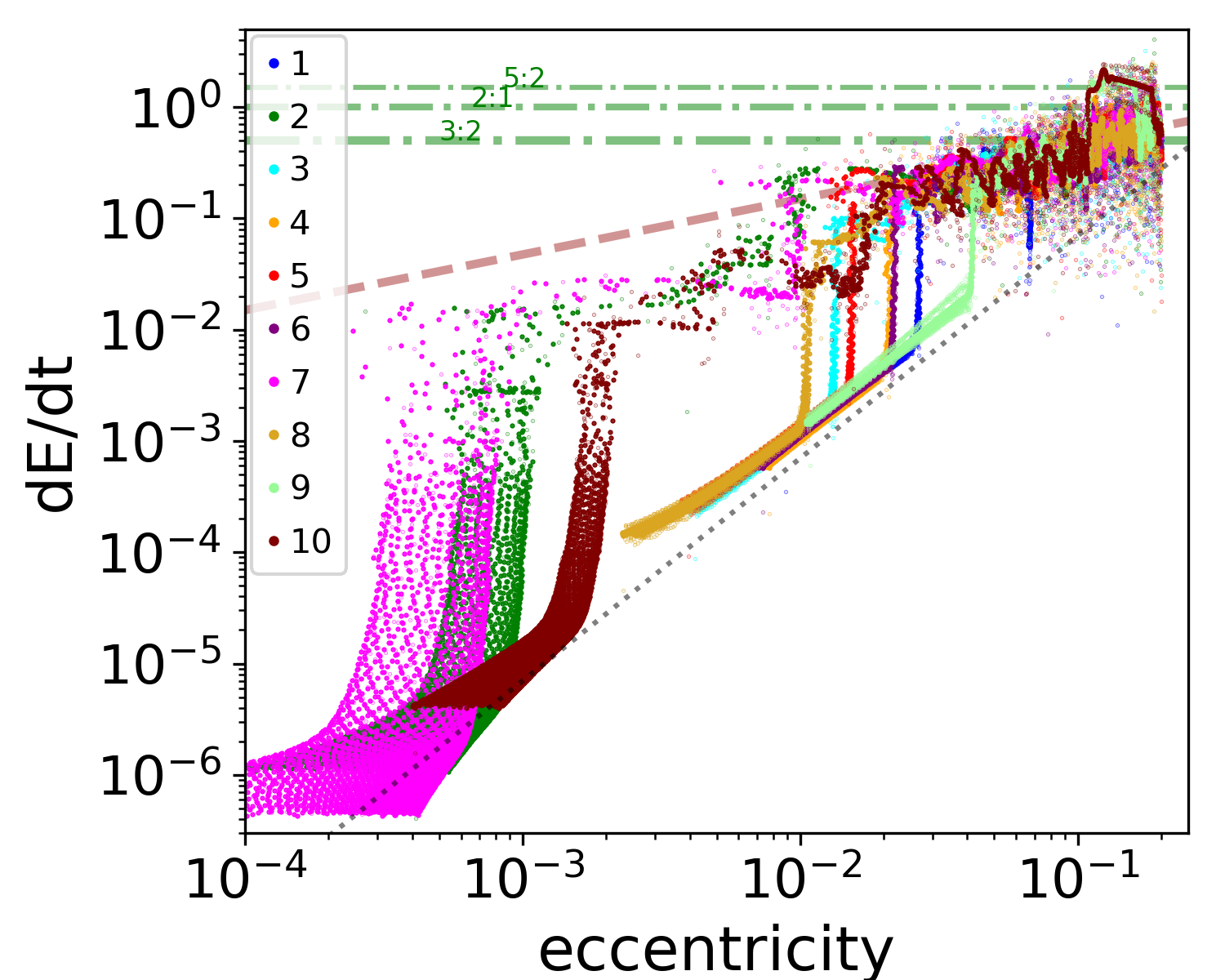}  
\includegraphics[trim=0 0 0 0,clip,width=3.2truein]{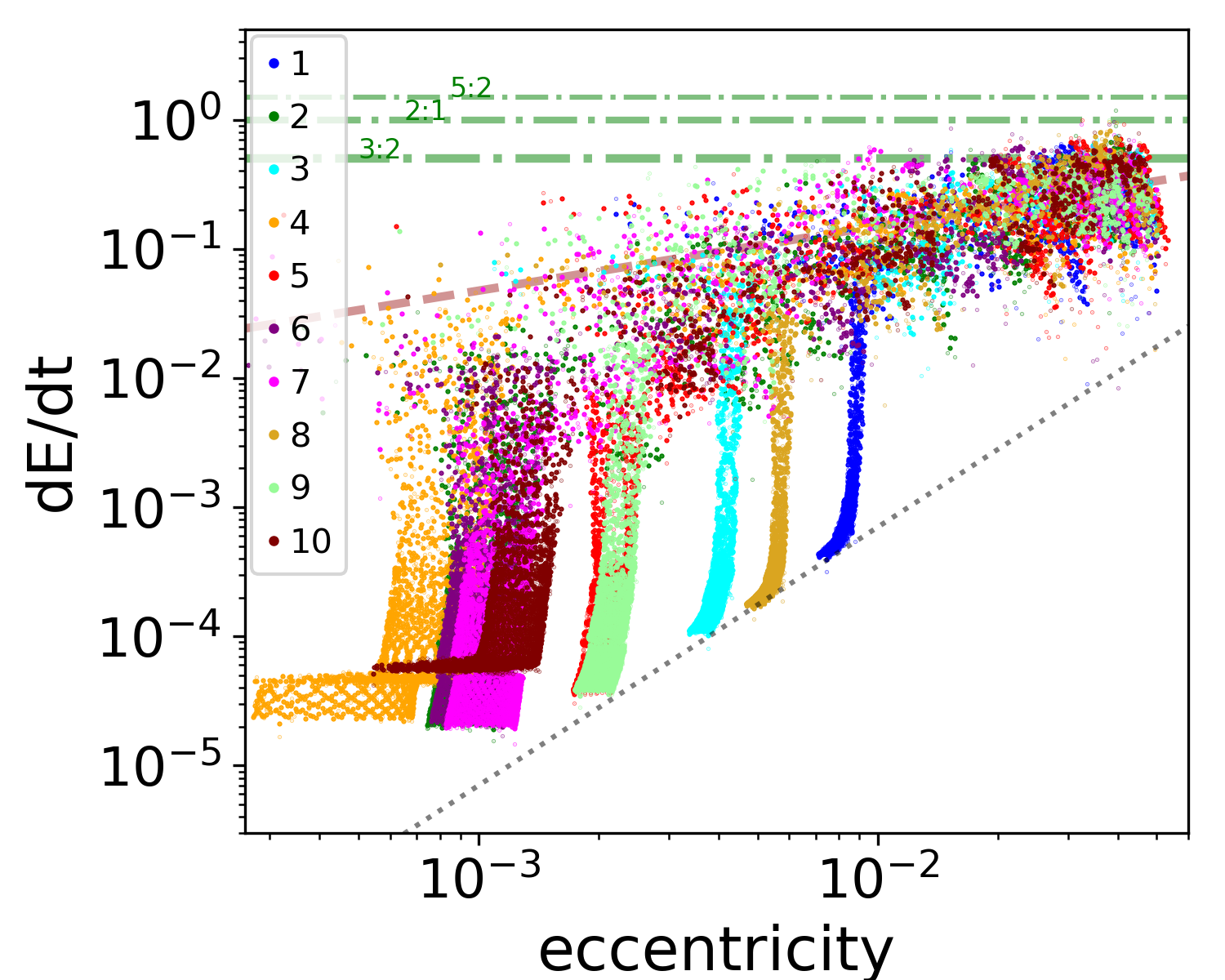}  
\caption{Energy dissipation rate vs eccentricity for the simulations shown in Figure \ref{fig:Ph_all}.  
Similar to Figure \ref{fig:Ee_De}
a) (on the left) The Ph$_F$ simulations. 
b) (on right) The Ph$_S$ simulations.  
\label{fig:Ee_Ph}} 
\end{figure*}

In the previous section (section \ref{sec:res}) orbital migration was forced to mimic drift caused
by tides dissipated in Mars.  Dissipation within the spinning body primarily served to damp
the body into the tidally locked state and keep it there.  However, in this section orbital migration and eccentricity damping arise solely due to satellite tides and  internal dissipation coming from damping in the springs.

We have run two series of simulations of a spinning resolved body, with axis ratios similar to Deimos and two series with axis ratios similar to Phobos.   The spinning body 
is alone   in orbit about a  point mass.  The  De$_F$ and Ph$_F$ series have initial eccentricity $e=0.2$ whereas the De$_S$ and Ph$_S$ series start with $e=0.05$.  The De$_F$ and Ph$_F$ series have 10 times higher damping in the springs (the $F$ represents fast)  than the  De$_S$ and Ph$_S$ series  (the $S$ represents slow).  The  De$_F$ and De$_S$ simulations of Deimos have the same node and spring network and the  Ph$_F$ and Ph$_S$ simulations of Phobos have the same network. The parameters for these simulations are listed in Tables  \ref{tab:common} and \ref{tab:ewob}.
 
By initially setting the body's spin above the mean motion and at moderate obliquity, 
we ensure that  the body begins to tumble soon afterwards.  Individual simulations 
in each series are begun at different spin, obliquities and mean anomaly.
In these simulations there is no forced migration $\tau_a^{-1} = 0$ and the quadrupole moment of the central body is zero.
As the energy is tidally dissipated, the eccentricity decreases and eventually the body drops into a spin synchronous state.
We measure the energy dissipation rates during the simulations and the eccentricities reached when the simulated elongated moon finally enters into the tidally locked state.  

As the tidal dissipation depends strongly on the ratio of semi-major axis to radius, tidal dissipation is stronger if we put the spinning body closer to the central mass.   
To reduce  simulation times, we reduced the central
mass $M_*$ to lower than that of Mars in units of the spinning body.
This is equivalent to inflating the radius of the spinning body  and it should not significantly affect the dynamics of the spin-orbit problem which is primarily sensitive to the body axis ratios \citep{quillen17_pluto}. The central mass $M_*$ and spring damping parameter $\gamma_s$ were adjusted to allow significant orbital eccentricity evolution during the simulations.  

The  De$_F$ and Ph$_F$ simulations were run to a maximum time of $t=10^6$ and
the De$_S$ and Ph$_S$ simulations to $2 \times 10^6$ in our gravitational units.  
For the De$_F$ and Ph$_F$ simulations, 
this maximum length of time is equivalent to about $5 \times 10^4$  orbital periods.  
As our mass-spring network is course and
randomly generated, the moments of inertia are not  the same as the triaxial boundary
used to confine the randomly generated  node positions.  We have included in Table \ref{tab:ewob} the asphericity parameter computed  from the nodes in the simulations.   
For the De$_F$ and De$_S$ simulations the asphericity parameter is similar to that of Deimos
and for the Ph$_F$ and Ph$_S$ simulations,  similar to that of Phobos.

Eccentricity and spin evolution for these series of simulations are shown
as a function of time in orbital periods in Figures  \ref{fig:De_all} and \ref{fig:Ph_all}.
We find that tumbling can last thousands of orbital periods, confirming the work by \citet{wisdom87}. 
Simulations with large semi-major axes  or lower levels of spring damping (the De$_S$ simulations
compared to the De$_F$ simulations) or stiffer springs spend even longer time in tumbling states, as expected. Prior to tidal lock, the simulations  primarily exhibited tumbling and only rarely experience short intervals  of spin-orbit resonance capture.   

Only the De$_F$ simulations show intervals captured into the 2:1 spin-orbit resonance (see Figure \ref{fig:De_all}a). The
De$_F$ simulations were begun at higher eccentricity than the De$_S$ simulations and the
spin reaches higher values at the beginning of the simulation.   
The De$_F$ simulations also have a higher spring damping parameter than the De$_S$ simulations. We ran a series of simulations at the 
lower damping rate but starting at $e=0.2$ and found that capture into the 2:1 resonance
can occur.   We suspect that the reason the De$_S$ simulations don't capture into the 2:1 spin-orbit 
resonance is not because these simulations have a lower dissipation rate, but
 because the mean spin values while tumbling are too low at lower eccentricity.  At lower 
 eccentricity, the tumbling body is  
 usually spinning slower than  the 2:1 spin-orbit resonance.     
For more elongated bodies, as seen in
the Ph$_F$ and Ph$_S$ simulations, capture into the 2:1 spin-orbit resonance was also unlikely,
though two of the Ph$_F$  simulations had intervals in the 3:1 spin-orbit resonance (see Figure \ref{fig:Ph_all}a).

The equation of motion for the classic conservative spin-orbit dynamical problem from Equation \ref{eqn:so1} can be written (following \citealt{celletti10,celletti14})
\begin{equation}
\ddot x + \frac{\alpha^2}{2}  V_x(x,t) = 0 \label{eqn:so2}
\end{equation}
with asphericity parameter $\alpha$ and 
\begin{equation}
V_x(x,t) \equiv \sum_{m \ne 0, m=-\infty}^\infty  W\left(\frac{m}{2}, e\right) \sin(2 x -  m t) .
\end{equation}
For $m=2,3,4$, the  coefficients are $W(1,e) = 1$, $W(3/2,e) =  \frac{7e}{2}  $  and 
$W(2,e) = \frac{17}{2} e^2$  \citep{cayley1859,goldreichpeale66} and
describe the  strengths of the 1:1, 3:2 and 2:1 spin-orbit resonances,  respectively.
Conditions for resonance overlap are computed using a distance between
spin-orbit resonances in units of the mean motion and resonant widths.
The resonant widths are 
 $\alpha$ for the 1:1 (spin synchronous) resonance,  $\alpha\sqrt{7e/2}$ for the 3:2 
spin-orbit resonance and $\alpha e\sqrt{17/2}$ for the 2:1 spin-orbit resonance.
The condition for overlap of the 1:1  and 3:2 spin-orbit resonances at low obliquity and eccentricity is
\begin{equation} 
\alpha  > \frac{1}{2 + \sqrt{14e}} \label{eqn:overlap} 
\end{equation}
 and  predicts when  chaotic tumbling takes place \citep{wisdom84}.

Equation \ref{eqn:overlap} implies that for asphericity parameter $\alpha > 1/2$ even a value of zero eccentricity would give chaotic behavior.    Asphericity parameters for Phobos and Deimos are both above 1/2. For $\alpha >1/2$, the width of the spin synchronous or 1:1 spin-orbit resonance exceeds the distance between the 1:1 and 3:2 spin-orbit resonances.   This means that the 3:2 spin-orbit resonance is diminished or swallowed up by the chaotic zone associated with the separatrix of the 1:1 resonance.  This would explain why we do not see
capture into the 3:2 spin-orbit resonance in these 4 series of simulations.     
We have run similar simulations
of rounder bodies, and those do spend time trapped in the 3:2 spin-orbit resonance.

A similarly derived condition for resonance overlap between 1:1 and 2:1 spin-orbit resonances
gives 
\begin{equation} 
 \alpha > \frac{1}{1 + e\sqrt{{17/}{2}} }. \label{eqn:overlap2}
 \end{equation}
For the asphericity of Phobos the overlap condition is satisfied at $e>0.08$
and for Deimos at $e>0.19$.  Our De$_F$  and De$_S$ simulations have asphericity parameter
similar to that of Deimos.   Contrary to expectation, the 2:1 resonance island
might increase with decreasing eccentricity rather than shrink as it emerges 
from the tumbling chaotic region.    This may explain why we tended to see
capture into the 2:1 resonance after the eccentricity had time to drop in the De$_F$ simulations
and why similar simulations of more elongated bodies (the Ph$_F$ and Ph$_S$ simulations) are  unlikely to spend time intervals in the 2:1 spin-orbit resonance.
 
At the beginning of the simulations and at high eccentricities
the Ph$_F$  simulations (see Figure \ref{fig:Ph_all}a) have high spin values, sometimes even
above 1.    The spin in these simulations in gravitational units is 0.3 which is approximately equal to that of Phobos in its gravitational units.   A spin value of 1 is spinning so fast that 
 material could leave the surface.      If Phobos were ever at large eccentricity
 $e \gtrsim 0.15$ it would have tumbled and during tumbling it would have experienced episodes where material was spun off its surface.

\subsection{Dissipation rates}
\label{sec:diss}

In Figure \ref{fig:Ee_De} we show normalized energy dissipation rates 
versus orbital eccentricity for the De$_F$  and De$_S$ series of simulations 
and in Figure \ref{fig:Ee_Ph} for the  Ph$_F$ and Ph$_S$ series. 
We box averaged the energy dissipation over intervals $10^2$ long and these points are shown as solid points in the Figure.  Open points show measurements from the simulation output without average, though we only plotted 1 out of 10 points to reduce confusion.
Each color in these figures shows a different simulation and points show simulation measurements
at different times.  Labels in the key on these figures refer to their order in Figures \ref{fig:De_all}
and \ref{fig:Ph_all}.

In Figures \ref{fig:Ee_De} an \ref{fig:Ee_Ph},
as eccentricity decreases due to tidal dissipation, a simulation  begins on the upper right side in one of  and drifts to the lower left.  The dissipation rate drops abruptly when the system falls into a spin synchronous state.   After reaching the spin synchronous resonance, obliquity, 
libration and non-principal axis rotation are damped and the energy dissipation rate can continue to drop.   Afterwards the dissipation rate is extremely low and so is the rate of eccentricity change.  
After the tidally locked state is reached, tidal evolution slows.
All the simulations end in the spin synchronous state, though some still exhibit non-principal axis rotation, non-zero obliquity and free libration amplitudes which take longer to decay. 

The energy tidally dissipated for a body that is spinning about a principal body axis
and in the spin synchronous state with little free libration is
\begin{align}
\dot E_{\rm ee} (e) &\approx - C_e n ( 7  e^2  + \sin^2 I)  \label{eqn:dEdt_ee}  \\
  C_e &\equiv  \frac{3}{2} M (na)^2 \frac{k_{2}}{Q} \left( \frac{R}{a} \right)^5 
                        \left(\frac{M_*}{M} \right)    \label{eqn:Ce}
\end{align}
\citep{kaula64,cassen80}, where $I$ is the obliquity.
Damping in eccentricity is accompanied by inward drift in orbital semi-major axis and tidal heating rates are extremely sensitive to semi-major axis. 
Equations \ref{eqn:dEdt_ee}, \ref{eqn:Ce} and $n\propto a^{-3/2}$ implies
that $\dot E_{\rm ee} \propto a^{-\frac{15}{2}}$.  
In Figures \ref{fig:Ee_De} and \ref{fig:Ee_Ph} we have corrected the 
energy dissipation rates by this factor of semi-major axis 
to take into account the slight decrease in semi-major axis occurring during the simulations.  
This correction slightly lifts the low eccentricity 
points on Figures \ref{fig:Ee_De} and \ref{fig:Ee_Ph}. 

The tidal energy dissipation rate for an object undergoing principal 
axis rotation and at zero obliquity that is spinning down (and not in a spin synchronous state)
\begin{align}
 \dot E_{\rm despin} 
 & \approx  - C_e |\omega - n| \label{eqn:dEdt_despin}
 \end{align} 
(e.g., \citealt{M+D} Equation 4.151 and 4.160) and using the coefficient $C_e$ defined in
Equation \ref{eqn:Ce}.
 This rate is expected to characterize the energy dissipation rate for tumbling states 
 \citep{wisdom87}.
By setting $\omega$ equal to the value expected in a  spin-orbit resonance,  Equation \ref{eqn:dEdt_despin}  gives an estimate for  energy dissipation rates in a  spin-orbit resonance.
For example in the 3:2 spin-orbit resonance set $\omega - n = 3/2 n -n = n/2$.
The mean tidal dissipation rate in spin-orbit resonance at low obliquity is approximately 
characterized by spin $\omega/n = m/2$ for $m\ne 2$,  and dissipation rate
\begin{equation}
\dot E_{m} (e) \approx  - C_e
 n  \left|\frac{m}{2} - 1\right| . \label{eqn:dEdt_m}
\end{equation}


The lower left hand side of Figure \ref{fig:Ee_De}a 
shows that simulations of a low obliquity, tidally locked body 
undergoing principal axis rotation show $dE/dt \propto e^2$, as given in Equation \ref{eqn:dEdt_ee}.   Because it is well studied, we  normalize the dissipation rates using the tidally locked states.
Checking to make sure the body is rotating about a principal axis (at low angle $J$), we measure the energy dissipation rate, obliquity and eccentricity in a simulation after it has reached a tidally locked state.  We use that state and Equation \ref{eqn:dEdt_ee} to estimate 
$C_e n$.  The energy dissipation rates are then divided by $C_e n$.  In the 2:1 spin-orbit resonance
$|\omega - n|/n = 1$ so our normalization  gives $|\dot E|/(C_e n) \approx 1$ in this resonance 
as long as the obliquity is low.  
The grey dotted lines in Figures \ref{fig:Ee_De} and \ref{fig:Ee_Ph} show $|\dot E|/(C_e n)  = 7 e^2$, 
consistent with Equation \ref{eqn:dEdt_ee} for a tidally locked state at $I=0$.
The minimum energy dissipation rates in the tidally locked states all lie near and parallel to this line, as expected.
Light green dot-dashed lines in Figure \ref{fig:Ee_De} and \ref{fig:Ee_Ph} show estimates for the 3:2, 2:1  and 5:2
spin-orbit resonances.   Four simulations 
in the De$_F$ series have intervals in the 2:1 spin-orbit resonance and while they are in that resonance, their energy dissipation rates lie near  the green dot-dashed line 
for that resonance in Figure \ref{fig:Ee_De}a.

Figures \ref{fig:Ee_De} and \ref{fig:Ee_Ph} show that the energy dissipation rate during tumbling
is 1 to 6 orders of magnitude above that in the tidally locked states, quantitatively confirming
the rough estimate by \citet{wisdom87}.
There is no analytical formula for the energy dissipation rate in a tumbling state,
though an approximate value was estimated by \citet{wisdom87} using Equation \ref{eqn:dEdt_despin} and replacing $|\omega - n| $ with $n$.
Figure \ref{fig:Ee_De}a shows that 
this is a good approximation as long as the eccentricity is above about 0.05.
Figures \ref{fig:Ee_De} and \ref{fig:Ee_Ph}  show that the energy dissipation rate is weakly dependent
on eccentricity with somewhat lower rates at lower eccentricities.
To roughly characterize the mean energy dissipation rate seen in the wobbling states
we plotted as a brown lines in Figures \ref{fig:Ee_De} and \ref{fig:Ee_Ph}
\begin{equation}
\dot E_{\rm wob} (e) = A_w C_e n e \label{eqn:Ae}
\end{equation}
using coefficient $A_w  = 14$ in Figure  \ref{fig:Ee_De}  and  $A_w  = 7$ in Figure  \ref{fig:Ee_Ph}.

We lack a proposed theoretical explanation for the dependence of 
dissipation rate while tumbling on eccentricity, but
the brown lines are a decent match to the energy dissipation
rates exhibited by the tumbling bodies.  In the classic spin-orbit  problem
(Equation \ref{eqn:so1}), the chaotic tumbling zone seen in Poincar\'e maps of the planar
system becomes increasingly narrow with decreasing eccentricity \citep{wisdom87}.   At low eccentricity, we would not expect
the average dissipation rate to be sensitive to eccentricity.  Equation \ref{eqn:so1} for the spin-orbit problem 
only allows a single angle for body rotation and tumbling bodies rotate
in three-dimensions.  For the three-dimensional problem chaotic regions are present
even at zero eccentricity \citep{wisdom87}.
 The  dependence on eccentricity is related to the size
of the configuration space that the tumbling body can explore in the three-dimensional  problem. 

With principal axis rotation, our previous work has
estimated how the tidal spin down rate depends on body shape (at most a factor of 2 difference
from the volume equivalent sphere; \citealt{quillen16_haumea}). 
However the energy dissipation rate in wobbling or tidally locked states probably does 
not depend on body shape in the same way.  The difference between the dissipation
rate  in the 2:1 spin orbit resonance and the green dashed line in Figure \ref{fig:Ee_De}a may be
due to this discrepancy or other degrees of freedom that have not damped away,  such as 
free libration, obliquity  and non-principal axis rotation. Similarly the differences in the estimated 
coefficients $A_w$ for the brown lines might be sensitive to body shape. The lines giving 
dissipation rates in spin-orbit resonances, tidally locked and tumbling states 
on Figures \ref{fig:Ee_De} and \ref{fig:Ee_Ph} could in future be corrected for a dependence on body shape. 

Our simulations, shown in Figures \ref{fig:Ee_De} and \ref{fig:Ee_Ph},  show that a tumbling body 
can remain tumbling long enough that it reaches a very low orbital eccentricity, 
 confirming the prediction by \citet{wisdom87}.  
If Deimos or Phobos were
excited into a tumbling state, they could end up at quite low eccentricity,   
so excitation of tumbling could  account for their current low eccentricities.
The eccentricity damping time for Deimos is currently of order $\sim 10^{12}$ years 
(as we listed in Table \ref{tab:phobos_deimos}; see section \ref{sec:phys}).
From comparing the height of the brown line and the grey dotted line at 
an eccentricity of 0.003, we see that the dissipation rate in the tumbling state
is about 5 orders of magnitude higher than that in the tidally locked state.
An increase in the dissipation rate by only 3 orders of magnitude could lower the eccentricity
damping timescale for Deimos 
enough to damp its eccentricity within the age of the Solar system (assuming $\mu Q = 10^{11}$ Pa).
With dissipation 5 orders of magnitude higher than in the tidally locked state,
Deimos would have had to be tumbling only for tens of millions of years before falling
into its current tidally locked rotation state in order for its eccentricity to have dropped
significantly due to tidal dissipation.  
 
We examine the tidal heat power that could be present in a tumbling state.  
Using Equation \ref{eqn:dEdt_m}, with $m=2$, representative of the tidal power in 
the tumbling state,
and using values for $k_2, Q$ discussed previously in section \ref{sec:phys} and listed in Table \ref{tab:phobos_deimos}, we estimate the tidal heating power
for Phobos and Deimos. We find $\dot E \approx 4.6 \times 10^5 $ W for Phobos and 
$\dot E \approx 5 $ W for Deimos.   Using a thermal conductivity of $K_T  = 1$ W/(m K) 
which is within a factor of a few similar to that  
of ice, rock or porous rock, we estimate a core to surface temperature difference
of $\Delta T = \dot E/ (K_T 4\pi R) = 3.2 $ K for Phobos.  These low powers and small
core to surface temperature difference imply that tidal heating is not likely to affect the composition 
 or interior structure  of Phobos or Deimos, even if they spent extended periods of time tumbling.
 
We consider the sensitivity of our estimates to rheology.
The tidal frequency $2(n-\omega) $ can be normalized by the viscoelastic relaxation time  $\tau_{\rm relax} \equiv \eta/\mu$ with $\eta$ the viscosity.  This gives a normalized tidal frequency  
${\bar\chi} = 2(n-\omega)\tau_{\rm relax} $.   We run our simulations in the regime with frequency  $\bar \chi  \ll 1$ where  the quality function is proportional to $\bar \chi$ for both Maxwell rheology and that exhibited by our
numerical simulations \citep{efroimsky13,frouard16}.   However, the short rotation periods
of Phobos and Deimos may put them in the opposite limit ($\bar \chi  \gg 1$) where the quality function 
is instead inversely proportional to ${\bar \chi}$ (again for both simulation and Maxwell rheologies).
In the tumbling state, the spin contains multiple frequency components, with average near the mean motion.   In the spin-synchronous state, the tidal perturbations are at multiples of the mean motion
with strongest term set by the mean motion.
We expect that the ratio of dissipation rates, between spin-synchronous and tumbling, would be similar for the high frequency limit as for the low frequency limit.  As a consequence,
the relative dissipation rates for the spin-synchronous and tumbling states for the 
low frequency limit, which is shown in our figures, can be used to
estimate  the relative  dissipation rates for bodies that are in the high frequency limit.
However, the higher spin rate spin-orbit resonances would have larger normalized tidal frequency $\bar \chi$ 
 than the 2:1 spin-orbit resonance and would  have lower dissipation rates for a Maxwell model, 
rather than higher ones as we estimated in Equation \ref{eqn:dEdt_m}.  
For the spin-orbit resonances, a Maxwell model would
 show different  relative dissipation rates. 
 
\subsection{Long-lived non-principal axis rotation}

The grey dotted lines  in Figures \ref{fig:Ee_De} and \ref{fig:Ee_Ph} representing the dissipation rates in the tidally locked states  are corrected for non-zero obliquity but not for long lasting non-principal axis rotation or free libration.  
The reason that points lie above the grey dotted line at $e\sim 0.01$ in Figure \ref{fig:Ee_Ph}a
showing the Ph$_F$ series is that the simulations have obliquity of a few degrees.  
\citet{frouard17} have estimated the sensitivity of the dissipation rate to free libration.  
Less straightforward is how the dissipation rate depends on non-principal axis rotation. 

\begin{figure}
\centering\includegraphics[trim=0 0 0 0,clip,width=3.2truein]{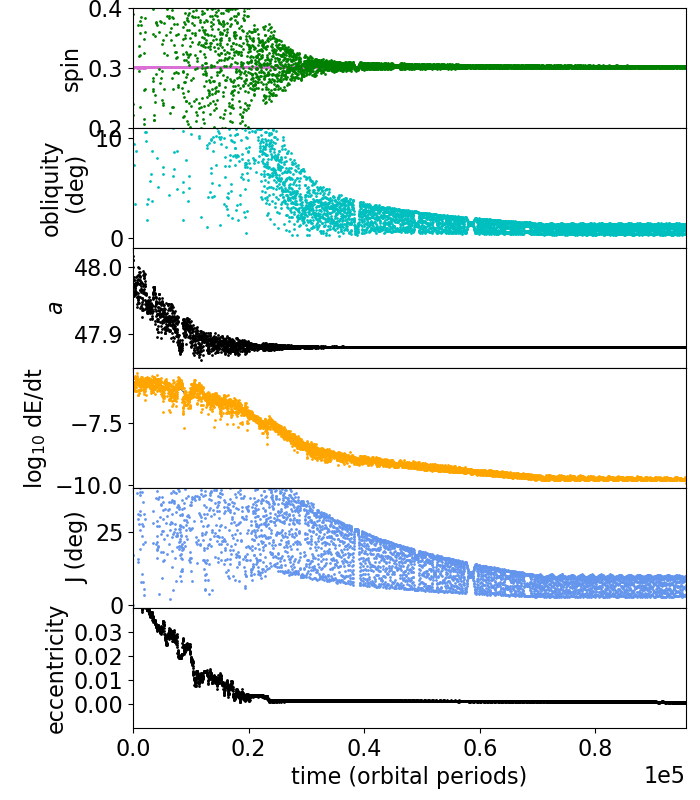} 
\caption{A simulation that displayed long lived non-principal axis rotation.
We show the 10th Ph$_S$ simulation with parameters
listed in Table \ref{tab:ewob} and also shown in Figures \ref{fig:Ph_all} and \ref{fig:Ee_Ph}.  The panels are similar to those of Figure \ref{fig:phobos21}.  
At $t \sim 7 \times 10^4$ the body enters
a long lived non-principal axis rotation state, evident because the non-principal
angle $J$ remains near $10^\circ$ and does not decrease (see blue points in the second
panel from the bottom).  
}
\label{fig:NPA}
\end{figure}

In some simulations we saw long lived non-principal axis rotation in the spin-synchronous state.
This is evident from looking at their oscillations in spin, eccentricity and from the non-zero  value of the non-principal angle $J$.
An example of long lived non-principal axis rotation is shown in Figure \ref{fig:NPA} which shows 
the 10-th simulation from the Ph$_S$  series. 
The panels in Figure \ref{fig:NPA} are similar to those in Figure \ref{fig:phobos21}.
In Figure \ref{fig:NPA}, the second panel from the bottom shows the non-principal angle $J$.  After $3 \times 10^4$  orbital periods,  the body has entered the spin synchronous state, but $J$ continues to be above zero. After $7 \times 10^4$ orbital periods, the mean $J$ stops decreasing and the body remains in a long lived non-principal axis rotation state.   
Many of the Ph$_S$ simulations show long lived non-principal axis rotation but with slowly decaying amplitudes.

The Ph$_S$ simulation shown in Figure \ref{fig:NPA} is striking because at later times there is no evidence for decay in the non-principal axis rotation and this suggests that the body has been trapped in a resonance with body axis precession frequencies.  Because the non-principal angle $J$ does not decay, the energy dissipation rate remains higher than it would be had the body damped into a state where it rotated about a principal body axis.
These resonances might have previously been seen as stable non-chaotic islands in a surface of section for the out-of-plane motion of a prolate, axisymmetric body in a circular orbit that exist
even at zero orbital eccentricity (see Figure 7
by \citealt{wisdom87}).  Such resonant islands probably exist for bodies that are not
axisymmetric.   Our Ph$_S$  simulations suggest that tidally dissipating bodies can be captured
into this type of resonance.   This might be more likely for longer and more slowly tidally
evolving bodies as we saw this behavior in the slower simulations of larger asphericity bodies. 

The energy dissipation rate is higher in the long lived non-principal axis rotation states than in a tidally locked state where the body is undergoing rotation about a principal body axis.
In the 4-th and 10-th Ph$_S$ simulations the energy dissipation rate flattens out at $\sim 5 \times 10^{-5}$ due to non-principal axis rotation.   The energy dissipation rate is about an order of magnitude above that expected for the tidally locked states. The discrepancy is noticeable because 
of the low eccentricity.  The other simulations in the Ph$_F$  series  also exhibit 
non-principal axis rotation but because the eccentricity is higher, the enhanced dissipation is less noticeable.

If the body stays captured in a resonance involving non-principal axis rotation
for a long time then the classic formula for tidal evolution in the spin synchronous
state would underestimate 
the energy dissipation rate and the associated eccentricity damping rate.
The size of the underestimate would depend on the amplitude of non-principal axis rotation,
 the eccentricity and obliquity.

\subsection{Chaotic Dynamics}

\citet{wisdom87} had speculated that once tumbling, eventually the body would exit the chaotic zone if it entered a sticky region near a stable resonant island.  
Our numerical study suggests that at low eccentricity, and for bodies with large asphericity such as Phobos and Deimos,  stability is only achieved in the spin synchronous state.  As the chaotic zone volume shrinks due to a slow decrease in eccentricity, a dissipating body can nevertheless 
remain for large periods of time in the tumbling state, even to eccentricities below 0.001.
In Figures \ref{fig:Ee_De} and \ref{fig:Ee_Ph},
the distributions of eccentricities reached when the simulations
drop into the spin-synchronous states  is fairly uniform and quite broad.  
We discuss possible qualitative explanations for this behavior. 

The classic spin-orbit dynamics problem (shown in Equations \ref{eqn:so1} and \ref{eqn:so2})  is that of a rigid non-spherical body undergoing principal axis rotation  that is orbiting a central point mass at zero obliquity.  The dynamics is conservative and described by a time dependent and periodic Hamiltonian that is a function of rotation angle and its time derivative \citep{celletti07,celletti10},   
\begin{equation}
H(p,x,t) = \frac{p^2}{2}  + \frac{\alpha^2}{2} V_x(x,t)
\end{equation}
with canonical momentum $p = \dot x$.
The Hamiltonian is a function of orbital eccentricity and the eccentricity is constant. 

The {\it dissipative spin-orbit problem}  includes in addition  a small tidal torque that is due to tidally generated heat,  giving equation of motion
\begin{equation}
\ddot x  + \frac{\alpha^2}{2} V_x(x,t) = -K_d ( L(e) \dot x - N(e))
\end{equation}
with unit-less dissipation factor $K_d$ defined in Equation \ref{eqn:Kd}. 
The functions $L(e), N(e)$ depend  on eccentricity  \citep{correia04,celletti10,celletti14}.
The  dissipative spin-orbit problem  also usually neglects variations in  eccentricity.   

For the De$_S$  and Ph$_S$ series of simulations we estimate dissipation parameter $K_d \sim 10^{-5}$ using estimates for the quality function, valid for the limit of low tidal frequency times viscoelastic relaxation time 
($\bar \chi<1$), that is based on the tidal frequency, spring strengths
and damping parameters ($k_2/Q \sim 0.038 {\bar \chi}/\mu$; for details see \citealt{frouard16}).  
The De$_F$ and Ph$_F$ simulations have 10 times higher dissipation rate than the De$_S$  and Ph$_S$ simulations.  
For the actual Phobos and Deimos,  $K_d \sim 10^{-7}$ and $10^{-9}$, respectively,  (see Table \ref{tab:phobos_deimos}).
The numerical values are a few orders of magnitude larger than those of the actual moons.
However, 
our numerical simulations have unit-less dissipation factor $K_d \ll 1$,  as do Phobos and Deimos.


Because the tidal dissipation rate is low, the dynamics of the related conservative problem can help interpret the dynamics of the non-conservative system.  The dissipative spin-orbit problem exhibits different kinds of attractors, including periodic, quasi-periodic and strange attractors  \citep{celletti07,celletti07b,melnikov14}.    
Regions in phase space can be divided into neighborhoods of attraction.
Some regions may not contain rotational invariant tori (this is referred to as
non-existence of rotational invariant tori  by \citealt{celletti07b}).  These regions
can contain strange attractors so some orbits remain in them forever.

The problem we have studied here is more complex than the dissipative spin-orbit dynamical model as our body spins in three-dimensions and the tidal dissipation also causes slow variations in eccentricity.   
Nevertheless as the eccentricity damping rate is slow, the dynamics of the related low dimension dissipative
and non-dissipative (classic) spin-orbit problems  might help interpret  our simulations. 
A  difference between the classic spin-orbit problem and the full 3d but conservative spin dynamics problem 
is that  the tumbling chaotic zone can persist even at zero eccentricity \citet{wisdom87}
when rotation out of the plane is allowed, but vanishes at zero eccentricity in the classic low dimensional  spin-orbit problem.

Our simulations have faster eccentricity damping rates in units of the mean motion  than Phobos and Deimos because we have inflated the radii of the spinning bodies.  The eccentricity damping rate depends on  dimensionless parameter 
\begin{equation}
K_e \equiv  K_d \xi \left(\frac{R}{a}\right)^2
\end{equation}
(following Equations \ref{eqn:dEdt_ee}, \ref{eqn:Ce}, \ref{eqn:dEdt_m}, \ref{eqn:Ae}) 
with $\frac{de^2}{dt} n^{-1} \sim K_e$  for the tumbling states. 
The dimensionless parameter $K_e$ characterizes the rate of eccentricity decrease.
The two factors of $R/a$ imply that $K_e\ll K_d$ for our simulations of Phobos and Deimos (and as desired)
 but the Ph$_S$ simulations have $K_e$ about $10^3$ times that of  Phobos and the De$_S$ simulations have $K_e$ about $10^6$ times that of  Deimos.
 
The existence of attractors in the dissipative spin-orbit system  implies that 
without eccentricity damping the dissipative spin-orbit system probably can remain in a tumbling state forever.  
The dissipation itself does not necessarily cause the system to exit the chaotic zone, even though it causes volume in phase space to shrink. In contrast, we can consider the evolution of the classic spin-orbit problem, without any dissipation but with an adiabatically decreasing eccentricity.  This is a time dependent but conservative problem.   As the eccentricity decreases, the width of the resonances decrease.   A system initially in a chaotic zone could  drop out of it, not because of dissipation,
but because the volume in phase space that is covered by the chaotic region shrinks as the eccentricity drops. 

A tumbling body  continues to 
dissipate energy, giving an inward drift in orbital semi-major axis.
At extremely low eccentricity, angular momentum conservation
can no longer be maintained  by an accompanying decrease in orbital eccentricity.   \citet{wisdom87} postulated that
``if the chaotic tumbling did persist to the point of completely damping the eccentricity, the chaotic tumbling 
would then be damped at a somewhat faster rate through the secular damping of the rotational Jacobi integral''.
Here the Jacobi integral is the rotational kinetic energy minus the rotation angular momentum times the mean motion.

Our simulations do not exhibit a transition to a faster mode of damping at low eccentricity.
Rather the eccentricities reached when dropping into the spin synchronous state seem
fairly evenly distributed.
The eccentricity distribution seems to be related to
 a reduction in the accessible  tumbling chaotic zone volume 
 as the eccentricity drops.  We use the chaotic zone width estimated from the classic low dimensional 
 spin-orbit problem to estimate a dependence of chaotic zone volume on eccentricity, even
 though the low dimensional spin-orbit problem might poorly approximate the full 3D problem.
 
At low eccentricity, the 3:2 spin orbit resonance can be considered a weak perturbation
primarily causing chaotic behavior on the separatrix of the 1:1 spin orbit resonance \citep{wisdom87}.
The 3:2 resonance perturbation term in the classic spin-orbit Hamiltonian would be  stronger  than the other spin-orbit  perturbations.
The width of the chaotic zone in the classic spin-orbit problem depends on the strength of the 3:2 resonance or 
$W(3/2,e) \propto e$.    The frequency of the perturbation angle (of the 3:2 resonant angle) 
compared to that of libration
in the 1:1 resonance for order unity asphericity  is $\lambda =\alpha/2$ which is of order unity. 
The perturbation is neither in the adiabatic or non-adiabatic limits, however \citet{shev12} estimates that the width of the chaotic zone near the separatrix has energy proportional to the perturbation strength in both  limits.
This implies that the separatrix width in energy is proportional to eccentricity.
The width in units of $\dot x$ would be proportional to $\sqrt{e}$ giving a chaotic zone 
volume in 2-dimensional phase space ($\dot x, x$) that is also proportional to $\sqrt{e}$.
The weak dependence of chaotic zone width on eccentricity in the spin-orbit problem was noticed by \citet{wisdom87} who commented in his Figure 3 caption regarding the chaotic zone for Deimos 
`Even with such a small eccentricity, the chaotic zone is not microscopic.'

We consider the volume in 2-dimensional phase space lost per decade of eccentricity decrease.
We start with points evenly distributed in the chaotic zone at the separatrix with volume $\propto \sqrt{e}$.   With eccentricity dropping adiabatically, the number of points leaving the chaotic zone in a decade in eccentricity would only be $\sqrt{10} \sim 3$ times more than in a consecutive lower eccentricity decade.   
For example the number of points leaving the chaotic zone at the separatrix between
$e = 0.01$ and $ 0.001$ should be about three times that leaving between $e = 0.001$ and 0.0001.
Such a  weak dependence of  chaotic zone volume on eccentricity would be consistent 
with the broad distributions of eccentricities reached when the simulations
drop into the spin-synchronous states seen in Figures \ref{fig:Ee_De} and \ref{fig:Ee_Ph}.  
We attribute the tendency for simulations of an initially tumbling body to drop into spin-synchronous state at very low eccentricity to the insensitivity of the chaotic zone volume to eccentricity.



\section{Spin excitation by impacts}
\label{sec:coll}

Large but rare collisions can significantly excite the spins of asteroids 
with diameters less than 100 km \citep{harris79,farinella98,henych13}.  As a consequence rare but nearly catastrophic impacts could also have pushed Phobos and Deimos out of their spin synchronous states.
Impacts below the catastrophic disruption threshold cannot significantly change orbital eccentricity.
This follows because satellite spin angular momentum 
is negligible compared to satellite orbital angular momentum. 
However a moon that is put in a tumbling state would experience enhanced dissipation until it returns into the tidally locked state.   During this time, the moon's orbital eccentricity would be damped.  

To push a moon out of tidal lock, the change in its spin $\delta \omega$
need only be $\omega/4$ as this is half of the distance to the 3:2 and 1:2 spin-orbit resonances.  With a tangential impulse from a grazing impact, the spin angular momentum after impact is 
\begin{equation}
L \approx  I \omega + \beta R m_p v_{\rm impact},
\end{equation}
where $\omega$ is the moon's initial spin,
$I \sim \frac{2}{5} M R^2$ is the moon's moment of inertia, $M,R$ are the moon's mass and radius, $m_p$ is the projectile mass and $v_{\rm impact}$ is the relative velocity of projectile and target moon.
The dimensionless parameter $\beta$ describes the total momentum transfer. 
Setting a change in angular momentum to $I \omega/4 $ gives a minimum mass ratio 
\begin{equation}
\frac{m_p}{M} \sim \frac{1}{10} \frac{R \omega}{\beta v_{\rm impact}}.
\end{equation}
The spin rotational velocity at the surface of the moon  $R\omega $ is 2.5 m/s for Phobos and 0.4 m/s for Deimos.  
We adopt a velocity of $v_{\rm impact}= 10$ km/s typical of collisions onto Mars \citep{neukem76,ivanov01}.
This gives a minimum projectile to target moon mass ratio of  order $2.5 \times 10^{-5} $ corresponding to 
a projectile diameter of about 0.7 km onto Phobos and approximately consistent with the models by \citet{ramsley19} for the projectile that formed the Stickney crater on Phobos.   
To kick Deimos out of the tidally locked state, a projectile to target moon mass ratio mass ratio of $4 \times 10^{-6}$ is required, corresponding to a 0.2 km diameter projectile.  Deimos is spinning less quickly than Phobos, so a lower mass ratio projectile can knock it out of the spin synchronous spin state.
These mass ratios lie in the regime of rare and nearly catastrophic collisions that can excite tumbling (see Figure 12 by \citealt{henych13}). 

How rare are such impacts?  The cross sectional areas of Phobos
and Deimos are approximately 400 km$^2$ and 120 km$^2$.
The mean time between impacts as a function of impactor diameter has been computed for Mars crossing asteroid 433 Eros (with area 1125 km$^2$) by \citet{richardson05} based on the asteroid size distribution model by \citet{obrien05}. We  estimate the mean time between impactors by multiplying by the ratio of cross sections.  This gives a few times $10^9$ years between
0.7 km diameter impactors on Phobos and $10^9$ years between 0.2 km diameter impacts
on Deimos.  These times neglect secondary impacts from Mars and that the bombardment rate
was higher in the past.  We conclude that as true for asteroids, 
rare and sub-catastrophic impacts capable of excitation of wobble  
 would also have occurred on Phobos and Deimos.

\section{Discussion and Summary}

Using a mass-spring model for modeling spin and tidally generated dissipation within non-spherical elastic bodies, we have explored scenarios for spin excitation of Phobos and Deimos. 
We test the possibility that crossing the strongest orbital resonances
(due to inwards drift of Phobos)
could have excited the spins of Phobos and Deimos. Our simulations of an initially tidally locked
Phobos crossing the 2:1 resonance
with Mars' figure rotation and an initially tidally locked Deimos crossing the 2:1 mean motion resonance with Phobos failed to show excitation of libration or wobbling in Phobos or Deimos. 
  
We have run longer simulations
of 2 drifting moons in orbit around Mars and also did not see tumbling excitation or large obliquity variations.    However, weak resonances can only be seen in extremely long simulations with slow drift rates  so longer simulations of more slowly drifting systems may yet reveal new resonant phenomena.
We have not yet simulated secular phenomena such as the secular resonances that were explored by \citet{yoder82}.   These would require longer integrations than run here and driving orbital precession frequencies or including the Sun and additional planets in the simulation.   Due to their elongated shape, body axis precession and libration frequencies for Phobos and Deimos are not slow (as they are for nearly spherical bodies like the Moon and Mercury), so strong coupling via secular spin resonances would be unexpected.  Our study rules out spin excitation by two  of
the strongest orbital resonances, however we do not necessarily rule out all types of spin resonances. 

We have carried out simulations of initially tumbling bodies in eccentric orbit
for bodies with axis ratios similar to Phobos and Deimos.
Our simulations dissipate energy within the simulated elastic body, so tidal
dissipation is directly tied to decreasing orbital semi-major axis and eccentricity.
Our simulations show that 
even at quite low eccentricity,  initially tumbling bodies remain so for thousands of orbital periods,
confirming prior results by \citet{wisdom87}.
When they finally fall into the spin synchronous state, the eccentricity can 
be substantially reduced from its initial value due to enhanced tidal dissipation while 
in the tumbling state.   Depending upon the eccentricity, energy dissipation rates
can be 1 to 6 orders of magnitude higher  when tumbling than when tidally locked.
Using our simulations we measure the mean dissipation 
rate in the tumbling state and find that it is weakly dependent on eccentricity.

We suspect that tidal dissipation does not necessarily cause the spinning body to leave the separatrix region or tumbling chaotic zone  because this region contains attractors.  From the conservative spin-orbit problem with one rotation angle at low eccentricity, we estimate that the volume of the chaotic zone is proportional to the square root of eccentricity.  With eccentricity adiabatically dropping, the likelihood that a body leaves the separatrix region can be estimated from the rate of chaotic zone volume decrease.
We attribute the tendency for simulations of an initially tumbling body to drop into spin-synchronous state at very low eccentricity to the insensitivity of the chaotic zone volume to eccentricity.
These rough arguments were made using qualitative properties of the classic and dissipative spin-orbit dynamical problems (depending upon only a single rotation angle) and could be tested and extended to the full three dimensional spin dynamics problem.

Because of the higher tidal dissipation rates when tumbling, we support the low
eccentricity scenario for Phobos proposed by \citet{yoder82}.
Most likely Phobos has remained at low eccentricity
during much of the age of the solar system.    If Phobos had been at higher eccentricity,
it would have spent time tumbling and it would have experienced spin out events.
Its tidal evolution rate in semi-major axis would have been too fast
to remain stable during the age of the Solar system.   Impact followed by tumbling can
decrease eccentricity, but would not increase it.   Consequently orbital resonance
crossing is still required to account for Phobos' non-zero eccentricity \citep{yoder82}.
As we support a low eccentricity history for Phobos, we prefer scenarios
where Phobos and Deimos  formed in or from a circumplanetary disk \citep{goldreich65,burns92,rosenblatt11,craddock11,rosenblatt16},
rather than were captured from heliocentric orbit \citep{burns92,lambeck79}.
In the scenario explored by 
\citet{rosenblatt16} for Phobos' and Deimos' formation from a circumplanetary disk, 
larger moons migrated outward due to interactions with a circumplanetary disk.   Interaction with
the disk  could also have damped orbital eccentricities, leaving them at low eccentricity
when the circumplanetary disk dissipated and prior to long timescale tidal evolution.
 
 Using the dissipation rates measured from our simulations,
we find that an episode of tumbling could have increased the dissipation rate 
for Deimos sufficiently that
its eccentricity could have been damped during the age of the Solar system.  This is despite its current slow
rate of tidal evolution. 
We estimate that both Deimos and Phobos likely experienced one or more sub-catastrophic impacts that would
have caused them to tumble.  As previously suggested  by \citet{wisdom87},
Deimos' low eccentricity may in part be due to an impact excited epoch of tumbling
that then lowered its eccentricity.  
The formation of the Stickney crater might have excited
tumbling on Phobos \citep{weidenschilling79,wisdom87,ramsley19}.  
Because  eccentricities can be substantially damped
while in a long lived tumbling state, spin excitation via rare impacts is
a  mechanism that can reduce satellite orbital eccentricity.    

We see some interesting phenomena in our simulations.  We find
that a tumbling body, after falling into the spin synchronous state, can 
spend time in a non-principal axis rotation state.  These probably
involve out of plane precession resonances.  These states  exhibit
enhanced dissipation which is particularly noticeable at low orbital eccentricity. 
In some simulations we have also seen resonant excitation of free libration that can also be long lived.
These types of excitation could cause enhanced tidal dissipation compared to 
that commonly predicted for a tidally locked low free libration, low obliquity body undergoing principal axis rotation.

In this paper we have focused on simulations of bodies with the axis ratios similar to  Phobos and Deimos.   With such extreme axis ratios, we primarily see tumbling or spin synchronous states.   However, simulations of initially tumbling somewhat rounder bodies show that they can more easily be captured into and spend more time in spin-orbit resonances such as the 3:2 or 2:1 resonances. 
Except for a short discussion at the end of section \ref{sec:diss}, 
sensitivity to rheology was neglected in this study.  This is because our simulations currently only exhibit a Kelvin-Voigt rheology and because we required long simulations, we compromised by poorly resolving the spinning bodies with our mass-spring model.   Sensitivity to rheology could be explored  in future studies using simulations of more accurately resolved bodies.

\vskip 1 truein 
Acknowledgements.

This material is based upon work supported in part by NASA grant 80NSSC17K0771, and National Science Foundation Grant No. PHY-1757062.
We acknowledge support from the NASA  Summer Undergraduate Program for Planetary Research (SUPPR) internship program.
We thank Juliana South, Tyler LaBree, John Siu, and Jonas Smucker for helpful discussions.
We are grateful to Christoph Lhotka,  Michael Efroimsky, Valeri Makarov, and James Kwiecinski for suggestions that inspired and significantly improved this manuscript.

\vskip 2 truein

{\bf Bibliography}

\bibliographystyle{elsarticle-harv}
\bibliography{refs_tides}

\end{document}